\documentclass[preprint,12pt]{elsarticle}

\usepackage{amssymb, amsmath, amsthm, amssymb, amsfonts, dsfont, mathrsfs, dirtytalk, tikz-cd, adjustbox, url, nicefrac, booktabs,array,longtable, tabu,bm, hyperref, graphicx, geometry,comment}
\usepackage{caption}

\usepackage[T1]{fontenc}    
\usepackage[english]{babel}
\usepackage[capitalise]{cleveref}

\renewcommand{\H}{\operatorname{H}}

\journal{XXX}

\begin{document}

\begin{frontmatter}



\title{The free energy principle made simpler but not too simple}


\author[1]{Karl Friston}
\author[1,2]{Lancelot Da Costa\corref{cor1}}
\ead{l.da-costa@imperial.ac.uk}
\author[1]{Noor Sajid}
\author[3,4,5]{Conor Heins}
\author[1,6]{Kai Ueltzhöffer}
\author[2]{Grigorios A. Pavliotis}
\author[1]{Thomas Parr}
\cortext[cor1]{Corresponding author}
\address[1]{Wellcome Centre for Human Neuroimaging, University College London, London, WC1N 3AR, United Kingdom}
\address[2]{Department of Mathematics, Imperial College London, London, SW7 2AZ, United Kingdom}
\address[3]{Department of Collective Behaviour, Max Planck Institute of Animal Behavior, Konstanz D-78457, Germany}
\address[4]{Centre for the Advanced Study of Collective Behaviour, University of Konstanz, D-78457, Germany}
\address[5]{Department of Biology, University of Konstanz, D-78457, Germany}
\address[6]{Department of General Psychiatry, Centre of Psychosocial Medicine, Heidelberg University, Voßstraße 2, D-69115 Heidelberg, Germany}



\begin{abstract}
This paper provides a concise description of the free energy principle, starting from a formulation of random dynamical systems in terms of a Langevin equation and ending with a Bayesian mechanics that can be read as a physics of sentience. It rehearses the key steps using standard results from statistical physics. These steps entail (i) establishing a particular partition of states based upon conditional independencies that inherit from sparsely coupled dynamics, (ii) unpacking the implications of this partition in terms of Bayesian inference and (iii) describing the paths of particular states with a variational principle of least action. Teleologically, the free energy principle offers a normative account of self-organisation in terms of optimal Bayesian design and decision-making, in the sense of maximising marginal likelihood or Bayesian model evidence. In summary, starting from a description of the world in terms of random dynamical systems, we end up with a description of self-organisation as sentient behaviour that can be interpreted as self-evidencing; namely, self-assembly, autopoiesis or active inference.
\end{abstract}

\begin{graphicalabstract}
\includegraphics[width=\textwidth]{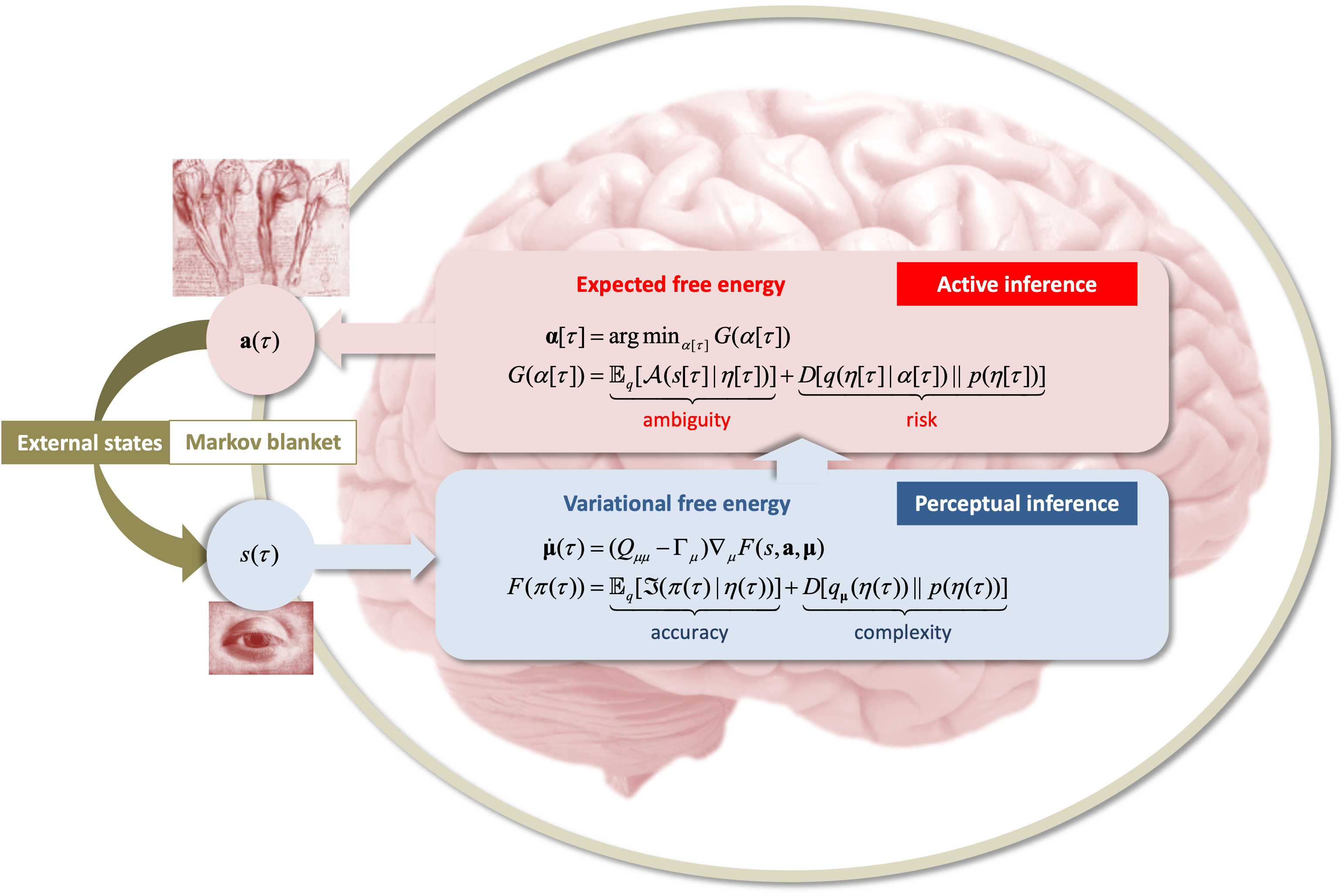}
\end{graphicalabstract}

\begin{highlights}
\item This paper provides a concise description of the free energy principle, starting from a formulation of random dynamical systems in terms of a Langevin equation and ending with a Bayesian mechanics that can be read as a physics of sentience.
\item Teleologically, the free energy principle offers a normative account of self-organisation in terms of optimal Bayesian design and decision-making, in the sense of maximising marginal likelihood or Bayesian model evidence.
\item In summary, starting from a description of the world in terms of random dynamical systems, we end up with a description of self-organisation as sentient behaviour that can be interpreted as self-evidencing; namely, self-assembly, autopoiesis or active inference.
\end{highlights}


\end{frontmatter}

\section{Introduction}
\label{sec: intro}

It is said that the free energy principle is difficult to understand. This is ironic on three counts. First, the free energy principle (FEP) is so simple that it is (almost) tautological. Indeed, philosophical accounts compare its explanandum to a desert landscape, in the sense of Quine~\cite{clarkWhateverNextPredictive2013}. Second, a tenet of the FEP is that everything must provide an accurate account of things that is as simple as possible—including itself. Finally, the FEP rests on straightforward results from statistical physics. This review tries to present the free energy principle as simply as possible but without sacrificing too much technical detail. It steps through the formal arguments that lead from a description of the world as a random dynamical system~\cite{crauelAttractorsRandomDynamical1994,arnoldRandomDynamicalSystems1998} to a description of self-organisation in terms of active inference and self-evidencing~\cite{hohwySelfEvidencingBrain2016}. The evidence in question is Bayesian model evidence, which speaks to the Bayesian mechanics on offer~\cite{fristonFreeEnergyPrinciple2019a}. These mechanics have the same starting point as quantum, statistical and classical mechanics. The only difference is that careful attention is paid to the way that the internal states of something couple to its external states.

To make the following account accessible, we use a conversational style, explaining the meaning of key mathematical expressions intuitively. Accordingly, simplifying notation and assumptions are used to foreground the basic ideas. Before starting, it might help to clarify what the free energy principle is—and why it is useful. Many theories in the biological sciences are answers to the question: “what must things do, in order to exist?” The FEP turns this question on its head and asks: “if things exist, what must they do?” More formally, if we can define what it means to be something, can we identify the physics or dynamics that a thing must possess? To answer this question, the FEP calls on some mathematical truisms that follow from each other. Much like Hamilton's principle of least action\footnote{Perhaps a better analogy would be Noether's theorem (Beren Millidge – personal communication)~\cite{noetherInvariantenBeliebigerDifferentialausdrucke1918}.}, it is not a falsifiable theory about the way ‘things’ behave—it is a general description of ‘things’ that are defined in a particular way. As such, the FEP is not falsifiable as a mathematical statement, but it may as well be falsifiable to the extent that its postulates refer to a specific class of empirical phenomena that the principle aims to describe.

Is such a description useful? In itself, the answer is probably no—in the sense that the principle of least action does not tell you how to throw a ball. However, the principle of least action furnishes everything we need to know to simulate the trajectory of a ball in a particular instance. In the same sense, the FEP allows one to simulate and predict the sentient behaviour of a particle, person, artefact or agent (i.e., some ‘thing’). This allows one to build sentient artefacts or use simulations as observation models of particles (or people). These simulations rest upon specifying a \textit{generative model} that is apt to describe the behaviour of the particle (or person) at hand. At this point, committing to a specific generative model can be taken as a commitment to a specific—and falsifiable—theory. Later, we will see some examples of these simulations.

The remaining sections describe the FEP. Each section focuses on an equation—or set of equations—used in subsequent sections. The ensuing narrative is meant to be concise, taking us from the beginning to the end as succinctly as possible. To avoid disrupting the narrative, we use footnotes to address questions that are commonly asked at each step. We also use figure legends to supplement the narrative with examples from neurobiology. Most of the following can be found in the literature~\cite{fristonFreeEnergyPrinciple2019a,dacostaBayesianMechanicsStationary2021a,fristonStochasticChaosMarkov2021}; however, there are a few simplifications that replace earlier accounts. 

\section{Systems, states and fluctuations}

We start by describing the world with a stochastic differential equation~\cite{pavliotisStochasticProcessesApplications2014}. So why start here? The principal reason is that we want a description that is consistent with physics. This follows because things like the Schrödinger equation in quantum mechanics, fluctuation theorems in statistical mechanics and the Lagrangian formulation of classical mechanics can all be derived from this starting point~\cite{seifertStochasticThermodynamicsFluctuation2012}. In short, if one wants a physics of sentience, this is the place to start. 

We are interested in systems that have characteristic states. Technically, this means the system has a pullback attractor; namely, a set of states a system will come to occupy from any initial state~\cite{crauelAttractorsRandomDynamical1994,crauelGlobalRandomAttractors1999}. Such systems can be described with stochastic differential equations, such as the Langevin equation describing the rate of change of some states $x(\tau)$, in terms of their flow $f(x)$, and random fluctuations $\omega(\tau)$. The fluctuations are usually assumed to be a normally distributed (white noise) process, with a covariance of $2\Gamma$:
\begin{equation}
\label{eq: system}
\begin{split}
\dot{x}(\tau) &=f(x)+\omega(\tau) \\
p(\omega \mid x) &=\mathcal{N}(\omega; 0,2 \Gamma) \Rightarrow p(\dot{x} \mid x)=\mathcal{N}(\dot{x}; f, 2 \Gamma) \\
p(x) &=?
\end{split}
\end{equation}
The dot notation denotes a derivative with respect to time\footnote{\textbf{Question}: why is the flow in \eqref{eq: system} not a function of time? Many treatments of stochastic thermodynamics allow for time-dependent flows when coupling one system (e.g., an idealised gas) to another (e.g., a heat reservoir), where it is assumed that the other system changes very slowly, e.g.,~\cite{jarzynskiNonequilibriumEqualityFree1997,seifertStochasticThermodynamicsFluctuation2012}. However, the ambition of the FEP is to describe this coupling under a partition of states. In this setting, separation of temporal scales is an emergent property, where \eqref{eq: system} holds at any given temporal scale. See~\cite{fristonFreeEnergyPrinciple2019a} for a treatment using the apparatus of the renormalisation group.}. This means that time and causality are baked into everything that follows, in the sense that states cause their motion. The Langevin equation is itself an approximation to a simpler mapping from some variables to changes in those variables with time. This follows from the separation into states and random fluctuations implicit in \eqref{eq: system}, where states change slowly in relation to fast fluctuations. This (adiabatic) approximation is ubiquitous in physics~\cite{carrApplicationsCentreManifold1982,hakenSynergeticsIntroductionNonequilibrium1978,koidePerturbativeExpansionIrreversible2017}. In brief, it means we can ignore temporal correlations in the fast fluctuations and assume—by the central limit theorem—that they have a Gaussian distribution. This equips the fluctuations with a probability density, which means we know their statistical behaviour but not their trajectory or path, which itself is a random variable~\cite{crauelAttractorsRandomDynamical1994,arnoldRandomDynamicalSystems1998,pavliotisStochasticProcessesApplications2014}.

The next step, shared by all physics, is to ask whether anything can be said about the probability density over the states—the ‘?’ in \eqref{eq: system}. A lot can be said about this probability density, which can be expressed in two complementary ways; namely, as \textit{density dynamics} using the Fokker-Planck equation (a.k.a. the forward Kolmogorov equation) or in terms of the probability of a path through state-space using the \textit{path-integral formulation}. The Fokker-Planck equation describes the change in the density due to random fluctuations and the flow of states through state-space~\cite{riskenFokkerPlanckEquationMethods1996,pavliotisStochasticProcessesApplications2014}:

\begin{equation}
\label{eq: FP eq}
\dot{p}(x, \tau)=\nabla \cdot(\Gamma \nabla-f(x)) p(x, \tau)
\end{equation}

The Fokker-Planck equation describes our stochastic process in terms of deterministic density dynamics—instead of specific realisations—where the density in question is over \emph{states}
$x(\tau)=x_{\tau}$. Conversely, the path-integral formulation considers the probability of a trajectory or \textit{path} $
x[\tau]\triangleq [x(t): 0 \leq t \leq \tau]
$ in terms of its \emph{action} $\mathcal{A}$ (omitting additive constants here and throughout)\footnote{\textbf{Question}: where does the divergence in the third equality come from? This term arises from the implicit use of Stratonovich path integrals~\cite{seifertStochasticThermodynamicsFluctuation2012}. Note that we have assumed that the amplitude of random fluctuations is state—and therefore path—independent in \eqref{eq: system}, which means we can place it outside the integral in the second equality.}:

\begin{equation}
\label{eq: action}
\begin{aligned}
\mathcal{A}(x[\tau]) &=-\ln p\left(x[\tau] \mid x_{0}\right) \\
&=\frac{\tau}{2} \ln |(4 \pi)^{n}\Gamma|+\int_{0}^{\tau} d t \mathcal{L}(x, \dot{x}) \\
\mathcal{L}(x, \dot{x}) &=\frac{1}{2}\left[(\dot{x}-f) \cdot \frac{1}{2 \Gamma}(\dot{x}-f)+\nabla \cdot f\right]
\end{aligned}
\end{equation}
 
Both the Fokker-Planck and path-integral formulations inherit their functional form from assumptions about the statistics of random fluctuations in \eqref{eq: system}. For example, the most likely path—or path of least action—is the path taken when the fluctuations take their most likely value of zero. This means that variations away from this path always increase the action. This is expressed mathematically by saying that its variation is zero when the action is minimised.
\footnote{Omitting the contribution of the divergence term in the Lagrangian to obtain the expression for the path of least action for simplicity, cf. \cite{durrOnsagerMachlupFunctionLagrangian1978}. Taking this simplification at face value means that we are either: 1) considering a description on a short time-scale as the flow can be approximated by a linear function with impunity (e.g., linear response theory, see \cite{pavliotisStochasticProcessesApplications2014}); or 2) we are considering the limit where random fluctuations have vanishingly small amplitude (e.g., precise particles, see Sections \ref{sec: precise} and \ref{sec: curious}).} 

\begin{equation}
\label{eq: path of least action}
\begin{aligned}
\mathbf{x}[\tau] &=\arg \min _{x[\tau]} \mathcal{A}(x[\tau]) \\
& \Leftrightarrow \delta_{x} \mathcal{A}(\mathbf{x}[\tau])=0 \\
& \Leftrightarrow \dot{\mathbf{x}}(\tau)=f(\mathbf{x})
\end{aligned}
\end{equation}
 								
In short, the motion on the path of least action is just the flow without random fluctuations. Paths of least action will figure prominently below; especially, when considering systems that behave in a precise or predictable way. We will denote the most likely states and paths with a bold typeface.

Although equivalent, the Fokker-Planck and path-integral formalisms provide complementary perspectives on dynamics. The former deals with time-dependent probability densities over \textit{states}, while the latter considers time-independent densities over \textit{paths}. The density over n states at any particular time is the time-marginal of the density over trajectories. These probabilities can be conveniently quantified in terms of their negative logarithms (or potentials) leading to surprisal and action, respectively (omitting the divergence of the flow in the last line for simplicity):

\begin{equation}
\label{eq: 5}
\begin{aligned}
\Im(x, \tau) &\triangleq -\ln p(x, \tau) \\
\mathcal{A}(x[\tau]) &\triangleq -\ln p\left(x[\tau] \mid x_{0}\right) \\
\H[p(x, \tau)] &=\mathbb{E}[\Im(x, \tau)] \\
\H\left[p\left(x[\tau] \mid x_{0}\right)\right] &=\mathbb{E}[\mathcal{A}(x[\tau])] \\
&=\frac{\tau}{2} \ln \left[(4 \pi)^{n}|\Gamma|\right]+\int_{0}^{\tau} d t \frac{1}{2} \mathbb{E}_{p(\omega)}\left[\omega(t) \cdot \frac{1}{2 \Gamma} \omega(t)\right]=\frac{\tau}{2} \ln \left[(4 \pi e)^{n}|\Gamma|\right]
\end{aligned}
\end{equation}

The second set of equalities shows that the uncertainty (or entropy) about states and their paths is the expected surprisal and action, respectively. Perhaps counterintuitively, the entropy of paths is easier to specify than the entropy of states. This follows because the only source of uncertainty about paths—given an initial state—are the random fluctuations~\cite{seifertStochasticThermodynamicsFluctuation2012,pavliotisStochasticProcessesApplications2014}, whose probability density does not change with time. The last pair of equalities in \eqref{eq: 5} show that the amplitude of random fluctuations determines the entropy of paths. Intuitively, if the fluctuations are large, then many distinct paths become equally plausible, and the entropy of paths increases\footnote{From a thermodynamic perspective, uncertainty about paths increases with temperature. For example, the Einstein-Smoluchowski relation relates the amplitude of random fluctuations to a mobility coefficient times the temperature $\Gamma=\mu_{m} k_{B} T$.}. 

\section{Solutions, steady-states and nonequilibria}

So far, we have equations that describe the relationship between the dynamics of a system and probability densities over fluctuations, states and their paths. This is sufficient to elaborate most physics. For example, we can use the Fokker-Planck or path-integral formalism to derive quantum mechanics, where the Fokker-Planck equation becomes the Schrödinger wave equation~\cite{arsenovicLagrangianFormSchrodinger2014}. We could focus on systems that comprise statistical ensembles of similar states to derive stochastic and statistical mechanics in terms of fluctuation theorems~\cite{seifertStochasticThermodynamicsFluctuation2012}. Finally, we could consider large systems—in which the fluctuations are averaged away—to derive classical mechanics such as electromagnetism and—with a suitable choice of potential functions—general relativity~\cite{krasnovGaugetheoreticApproachGravity2012,kleemanPathIntegralFormalism2015}. All of these mechanics require some boundary conditions: for example, a Schrödinger potential in quantum mechanics, a heat bath or reservoir in statistical mechanics and a classical potential for Lagrangian mechanics. At this point, the FEP steps back and asks, where do these boundary conditions come from? Indeed, this was implicit in Schrodinger's question:

\say{\textit{How can the events in space and time which take place within the spatial boundary of a living organism be accounted for by physics and chemistry?}}~\cite{schrodingerWhatLifeMind2012}.

We read a boundary in a statistical sense as a Markov boundary~\cite{pearlGraphicalModelsProbabilistic1998}\footnote{A Markov boundary is a subset of states of the system that renders the states of a ‘thing’ or particle conditionally independent from all other states~\cite{pearlCausality2009}.}. Why? Because the only thing we have at hand is a probabilistic description of the system. And the only way to separate the states of something from its boundary states is in terms of probabilistic independencies—in this instance, conditional independencies\footnote{Noting that if two subsets of states were independent, as opposed to being conditionally independent, we would be describing two separate systems.}. This means we need to identify a partition of states that assigns a subset to a ‘thing’ or particle and another subset to the boundary that separates the thing from some ‘thing’ else. In short, one has to define ‘thingness’ in terms of conditional independencies.

However, if things are defined in terms of conditional independencies and conditional independencies are attributes of a probability density, where does the density come from? The Fokker-Planck equation shows that the density over states depends upon time, even if the flow does not. This means that if we predicate ‘thingness’ on a probability density, it may only exist for a vanishingly small amount of time. This simple observation compels us to consider probability densities that do not change with time, namely: (i) steady-state solutions to the Fokker-Planck equation or (ii) the density over paths. We will start with the (slightly more delicate) treatment of steady-state solutions and then show that the (slightly more straightforward) treatment of densities over paths leads to the same notion of ‘thingness’.

The existence of things over a particular timescale implies the density in \eqref{eq: FP eq} does not change over that timescale. This is what is meant by a steady-state solution to the Fokker-Planck equation. The ensuing density is known as a \textit{steady-state density} and, in random dynamical systems, implies the existence of a pullback attractor~\cite{crauelAttractorsRandomDynamical1994,arnoldRandomDynamicalSystems1998}. The notion of an attractor is helpful here, in the sense that it comprises a set of characteristic states, to which the system is attracted over time\footnote{More precisely, the time-dependent solutions to the Fokker-Planck equation will tend towards the stationary solution, or steady-state. In other words, the steady-state density becomes a point attractor in the space of probability densities.}. In short, to talk about ‘things’, we are implicitly talking about a partition of states in a random dynamical system that has an attracting set—i.e., a steady-state solution to the Fokker-Planck equation. In short, we consider systems that self-organise towards a steady-state density\footnote{At this point, the formalism applies equally to steady-states with a high or low entropy, as we have not committed to a particular form of the steady-state density. Later, we will specialise to steady-states with a low entropy to characterise the sort of self-organisation that describes biological systems, e.g., swarming or flocking~\cite{hakenSynergeticsIntroductionNonequilibrium1978,nicolisSelforganizationNonequilibriumSystems1977}}. This solution is also known as a nonequilibrium steady-state (NESS) density, where the ‘nonequilibrium’ aspect rests upon solenoidal flow, as we will see next.

The existence of a solution to the Fokker-Planck equation—i.e., the existence of something—means that we can express the flow of states in terms of the steady-state density (or corresponding surprisal) using a generalisation of the Helmholtz decomposition. This decomposes the flow into conservative (rotational, divergence-free) and dissipative (irrotational, curl-free) components—with respect to the steady-state density—referred to as \textit{solenoidal} and \textit{gradient} flows, respectively~\cite{grahamCovariantFormulationNonequilibrium1977,eyinkHydrodynamicsFluctuationsOutside1996,shiRelationNewInterpretation2012,maCompleteRecipeStochastic2015,barpUnifyingCanonicalDescription2021,dacostaEntropyProductionStationary2022,pavliotisStochasticProcessesApplications2014}:

\begin{equation}
\label{eq: 6}
\begin{aligned}
&\dot{p}(x)=0 \Leftrightarrow f(x)=\Omega(x) \nabla \Im(x)-\Lambda(x)=\underbrace{Q(x) \nabla \Im(x)}_{\text {Solenoidal flow }}-\underbrace{\Gamma \nabla \Im(x)}_{\text {Gradient flow }}-\Lambda(x) \\
&\Im(x)=-\ln p(x), \quad Q=-Q^{T}, \quad 
\Lambda_{i}\triangleq\sum_{j} \frac{\partial \Omega_{i j}}{\partial x_{j}} = \sum_j \frac{\partial Q_{ij}}{\partial x_{j}}.
\end{aligned}
\end{equation}
  
This can be understood intuitively as a decomposition of the flow into two parts. The first (conservative) part of the flow is a \textit{solenoidal} circulation on the isocontours of the steady-state density (or surprisal). This component breaks \textit{detailed balance} and renders the steady-state density a \textit{nonequilibrium} steady-state density~\cite{aoPotentialStochasticDifferential2004,yuanPotentialFunctionDynamical2011}. The second (dissipative) part performs a (natural) gradient descent on the steady-state surprisal and depends upon the amplitude of random fluctuations~\cite{girolamiRiemannManifoldLangevin2011,amariNaturalGradientWorks1998}. The final term, $\Lambda$, can be regarded as a correction term, which is neither curl-free nor divergence-free, and which ensures that the probability density remains constant over time~\cite{dacostaEntropyProductionStationary2022}.

\subsection{Summary}

We now have a probabilistic description of a system in terms of a (NESS) density that admits conditional independencies among states. These conditional independencies are necessary to separate the states of things from their boundaries. In the next step, we will see how conditional independencies inherit from sparse coupling among states—and how they are used to establish a particular partition of states.

\section{Particles, partitions and things}
\label{sec: particles, partitions and things}

In associating some (stochastic differential) equations of motion with a unique (NESS) density, we have a somewhat special setup, in which the influences entailed by the equations of motion place constraints on the conditional independencies of the NESS density
. These conditional independencies can be used to identify a particular partition of states into \textit{external}, \textit{sensory}, \textit{active} and \textit{internal} states as summarised below. This is an important move because it separates the states of a \textit{particle} (i.e., internal states and their sensory and active states) from the remaining (i.e., \textit{external}) states. However, to do this we have to establish how the causal dynamics in \eqref{eq: system} underwrite conditional independencies. This can be done simply by using the curvature (Hessian) of surprisal as follows:
\begin{equation}
\label{eq: 7}
\begin{aligned}
\left(x_{u} \perp x_{v}\right) \mid b & \Leftrightarrow p(x)=p\left(x_{u} \mid b\right) p\left(x_{v} \mid b\right) p(b) \\
& \Leftrightarrow \Im(x)=\Im\left(x_{u} \mid b\right)+\Im\left(x_{v} \mid b\right)+\Im(b) \Leftrightarrow \frac{\partial^{2} \Im}{\partial x_{u} \partial x_{v}}=\mathbf{H}_{u v}=\mathbf{H}_{v u}=0.
\end{aligned}
\end{equation}
This says that if the $u$-th state is conditionally independent of the $v$-th state, given the remaining states $b$, then the corresponding element of the curvature—or Hessian matrix—of surprisal must be zero. Conversely, a zero entry in the Hessian implies conditional independence. In sum, any two states are conditionally independent if, and only if, the change of surprisal with one state does not depend on the other. We can now use the Helmholtz decomposition \eqref{eq: 6} to express the Jacobian—i.e., the (linear) coupling—of the flow in terms of the Hessian—that entails conditional independencies (with a slight abuse of the dot product notation):
\begin{equation}
\label{eq: 8}
\begin{aligned}
f(x) &=\Omega \nabla \Im-\Lambda \\
& \Rightarrow \mathbf{J}=\Omega \mathbf{H}+\nabla \Omega \cdot \nabla \Im-\nabla \Lambda \\
& \Rightarrow \mathbf{J}_{u v}=\frac{\partial f_{u}}{\partial x_{v}}=\sum_{i} \Omega_{u i} \mathbf{H}_{i v}+\sum_{i} \frac{\partial \Omega_{u i}}{\partial x_{v}} \frac{\partial \Im}{\partial x_{i}}-\sum_{i} \frac{\partial^{2} \Omega_{u i}}{\partial x_{i} \partial x_{v}}.
\end{aligned}
\end{equation}
We can now define \textit{sparse coupling} as a solution to this equation, in which all the terms are identically zero\footnote{This implicitly precludes edge cases, in which some non-zero terms cancel.}:
\begin{equation}
\label{eq: 9}
\left.\begin{array}{l}
Q_{u i} \mathbf{H}_{i v} \\
\Gamma_{u} \mathbf{H}_{u v} \\
\partial \Omega_{u i} / \partial x_{v}
\end{array}\right\}=0: \forall i \Rightarrow \mathbf{J}(x)_{u v}=0.
\end{equation}
Sparse coupling means that the Jacobian coupling states $u$ and $v$ is zero, i.e., an absence of coupling from one state to another. This definition precludes solenoidal coupling with $ u$ that depends on $v$. Because
$\mathbf{H}(x)_{v v}$ and $\Gamma_{u}$ are positive definite, sparse coupling requires associated elements of the solenoidal operator and Hessian to vanish at every point in state-space, which in turn, implies conditional independence:
\begin{equation}
\label{eq: 10}
\begin{aligned}
Q_{u v} \mathbf{H}_{v v} &=0 \Rightarrow Q_{u v}=-Q_{v u}=0 \\
\Gamma_{u} \mathbf{H}_{u v} &=0 \Rightarrow \mathbf{H}_{u v}=\mathbf{H}_{v u}=0 \Leftrightarrow\left(x_{u} \perp x_{v}\right) \mid b.
\end{aligned}
\end{equation}

In short, sparse coupling means that any two states are conditionally independent \textit{if one state does not influence the other}. This is an important observation; namely, that sparse coupling implies a NESS density with conditional independencies. In turn, this means any dynamical influence graph with absent or directed edges admits a \textit{Markov blanket} (the states $b$ above). These independencies can now be used to build a particular partition as follows:
\begin{itemize}
    \item The Markov boundary $a\subset x$  of a set of internal states $\mu \subset x$ is the minimal set of states for which there exists a nonzero Hessian submatrix: $\mathbf{H}_{a \mu} \neq 0$. In other words, the internal states are independent of the remaining states, when conditioned upon their Markov boundary, called \textit{active states}. The combination of active and internal states will be referred to as \textit{autonomous states}: $\alpha=(a, \mu)$.
    \item The Markov boundary $s \subset x$ of autonomous states is the minimal set of states for which there exists a nonzero Hessian submatrix: $\mathbf{H}_{s \alpha} \neq 0$. In other words, the autonomous states are independent of the remaining states, when conditioned upon their Markov boundary, called \textit{sensory states}. The combination of active and sensory (i.e., boundary) states constitute \textit{blanket states}: $b=(s, a)$. The internal and blanket states will be referred to as \textit{particular states}: $\pi=(s, \alpha)=(b, \mu)$.
    \item The remaining states constitute \textit{external states}: $x=(\eta, \pi)$. 
\end{itemize}
The names of active and sensory (i.e., blanket) states inherit from the literature, where they are often associated with biotic systems that act on—and sense—their external milieu\footnote{\textbf{Question}: why does a particular partition comprises four sets of states? In other words, why does a particular partition consider two Markov boundaries; namely, sensory and active states? The reason is that this is the minimal partition that allows for directed coupling with blanket states. For example, sensory states can influence internal states—and active states can influence external states—without destroying the conditional independencies of the particular partition (these directed influences are illustrated in the upper panel of Figure \ref{fig: 1} as dotted arrows).}. In this setting, one can regard external states as influencing internal states via sensory states (directly or through active states). And internal states influence external states via active states (directly or through sensory states\footnote{\textbf{Question}: does this mean that I can act on my world through my sense organs? Yes: much of biotic action is mediated by (active) motile cytoskeletal filaments, muscles and secretory organs that lie beneath (sensory) epithelia, such as receptors on the skin or a cell surface.}). We will see later how this implies a synchronisation between internal and external states, in the sense that internal states can be seen as actively inferring external states~\cite{dacostaBayesianMechanicsStationary2021a,fristonStochasticChaosMarkov2021}. The ensuing conditional independencies implied by a particular partition can be summarised as follows:
\begin{equation}
\label{eq: 11}
\begin{aligned}
\mathbf{J}_{\mu \eta} &=0 \Rightarrow \mathbf{H}_{\mu \eta}=0 \Leftrightarrow(\mu \perp \eta) \mid b \\
\mathbf{J}_{a \eta} &=0 \Rightarrow \mathbf{H}_{a \eta}=0 \Leftrightarrow(a \perp \eta) \mid s, \mu \\
\mathbf{J}_{s \mu} &=0 \Rightarrow \mathbf{H}_{s \mu}=0 \Leftrightarrow(s \perp \mu) \mid a, \eta
\end{aligned}
\end{equation}
 					
A normal form for the flow and Jacobian of a particular partition—with sparse coupling—can be expressed as follows, where  $\alpha=(a, \mu)$ and $\beta=(\eta, s)$:
\begin{equation}
\label{eq: 12}
\begin{split}
f(x)&=\Omega \nabla \Im-\Lambda\\
\left[\begin{array}{l}
f_{\eta}(\eta, b) \\
f_{s}(\eta, b) \\
f_{a}(b, \mu) \\
f_{\mu}(b, \mu)
\end{array}\right]&=\left[\begin{array}{cccc}
Q_{\eta \eta}-\Gamma_{\eta} & Q_{\eta s} & & \\
-Q_{\eta s}^{T} & Q_{s s}-\Gamma_{s} & & \\
& & Q_{a a}-\Gamma_{a} & Q_{a \mu} \\
& & -Q_{a \mu}^{T} & Q_{\mu \mu}-\Gamma_{\mu}
\end{array}\right]\left[\begin{array}{c}
\nabla_{\eta} \Im(\eta \mid b) \\
\nabla_{s} \Im(b \mid \eta) \\
\nabla_{a} \Im(b \mid \mu) \\
\nabla_{\mu} \Im(\mu \mid b)
\end{array}\right]-\Lambda\\
\mathbf{J}(x)&=\Omega \mathbf{H}+\nabla \Omega \cdot \nabla \Im-\nabla \Lambda\\
\left[\begin{array}{lllll}
\mathbf{J}_{\eta \eta \eta} & \mathbf{J}_{\eta s} & \mathbf{J}_{\eta a} & \\
\mathbf{J}_{s \eta} & \mathbf{J}_{s s} & \mathbf{J}_{s a} & \\
& \mathbf{J}_{a s} & \mathbf{J}_{a a} & \mathbf{J}_{a \mu} \\
& \mathbf{J}_{\mu s} & \mathbf{J}_{\mu a} & \mathbf{J}_{\mu \mu}
\end{array}\right]&=\left[\begin{array}{cccc}
Q_{\eta \eta}-\Gamma_{\eta} & Q_{\eta s} & & \\
-Q_{\eta s}^{T} & Q_{s s}-\Gamma_{s} & & \\
& & Q_{a a}-\Gamma_{a} & Q_{a \mu} \\
& & -Q_{a \mu}^{T} & Q_{\mu \mu}-\Gamma_{\mu}
\end{array}\right]\left[\begin{array}{cccc}
\mathbf{H}_{\eta \eta} & \mathbf{H}_{\eta s} & & \\
\mathbf{H}_{\eta s}^{T} & \mathbf{H}_{s s} & \mathbf{H}_{s a} & \\
& \mathbf{H}_{s a}^{T} & \mathbf{H}_{a a} & \mathbf{H}_{a \mu} \\
& & \mathbf{H}_{a \mu}^{T} & \mathbf{H}_{\mu \mu}
\end{array}\right]\\
&+\nabla \Omega \cdot \nabla \Im-\nabla \Lambda\\
\nabla_{\eta} \Omega_{\alpha \alpha}&=0, \quad \nabla_{\mu} \Omega_{\beta \beta}=0, \quad \nabla_{\eta} \Lambda_{\alpha}=0, \quad \nabla_{\mu} \Lambda_{\beta}=0
\end{split}
\end{equation}

This normal form means that particular partitions can be defined in terms of sparse coupling. Perhaps the simplest definition—that guarantees a Markov blanket\footnote{In the absence of solenoidal coupling between autonomous and non-autonomous states, and constraints on the partial derivatives of the solenoidal coupling in \eqref{eq: 12}; i.e., solenoidal coupling among autonomous states does not depend upon external states. Similarly, for non-autonomous and internal states.}—is as follows: \textit{external states only influence sensory states and internal states only influence active states}. This means that sensory states are not influenced by internal states and active states are not influenced by external states,
\begin{equation}
\label{eq: 13}
\left[\begin{array}{c}
\dot{\eta}(\tau) \\
\dot{s}(\tau) \\
\dot{a}(\tau) \\
\dot{\mu}(\tau)
\end{array}\right]=\left[\begin{array}{l}
f_{\eta}(\eta, s, a)+\omega_{\eta}(\tau) \\
f_{s}(\eta, s, a)+\omega_{s}(\tau) \\
f_{a}(s, a, \mu)+\omega_{a}(\tau) \\
f_{\mu}(s, a, \mu)+\omega_{\mu}(\tau)
\end{array}\right]
\end{equation}
and the noise processes $\omega_{i}(\tau), i \in \{\eta, s, a, \mu \}$ are independent. Under this sparse coupling, it is simple to show that not only are internal and external states conditionally independent, but their paths are conditionally independent, given initial states, using the path integral formulation.

The uncertainty (i.e., entropy) over paths derives from random fluctuations. This means that if we knew all the influences on the flow at every point in time, we can evaluate the entropy of external and internal paths from \eqref{eq: 5}:
\begin{equation}
\label{eq: 14}
\begin{aligned}
\H\left[p\left(\eta[\tau] \mid b[\tau], x_{0}\right)\right] &=\frac{\tau}{2} \ln \left[(4 \pi e)^{n_\eta}\left|\Gamma_{\eta}\right|\right] \\
\H\left[p\left(\mu[\tau] \mid b[\tau], x_{0}\right)\right] &=\frac{\tau}{2} \ln \left[(4 \pi e)^{n_\mu}\left|\Gamma_{\mu}\right|\right] \\
& \Rightarrow \\
\H\left[p\left(\eta[\tau] \mid b[\tau], x_{0}\right)\right] &=\H\left[p\left(\eta[\tau] \mid \mu[\tau], b[\tau], x_{0}\right)\right] \Rightarrow(\mu[\tau] \perp \eta[\tau]) \mid b[\tau], x_{0} \\
\H\left[p\left(\mu[\tau] \mid b[\tau], x_{0}\right)\right] &=\H\left[p\left(\mu[\tau] \mid \eta[\tau], b[\tau], x_{0}\right)\right] \Rightarrow(\mu[\tau] \perp \eta[\tau]) \mid b[\tau], x_{0}
\end{aligned}
\end{equation}

The final equalities say that the uncertainty about external (resp., internal) paths does not change when we know the internal (resp., external) path because external (resp., internal) states do not influence internal (resp., external) flow. This means the external and internal paths do not share any mutual information and are therefore independent when conditioned on blanket paths (and initial states). From \eqref{eq: 11}, the initial external and internal states are themselves independent, when conditioned on blanket states. 

Note that the conditional independence of paths inherits directly from the sparse coupling, without any reference to the NESS density or Helmholtz decomposition. This can be seen clearly by replacing the partial derivatives in \eqref{eq: 7} with functional derivatives and noting, from \eqref{eq: 12}, that there are no flows that depend on both internal and external states:
\begin{equation}
\label{eq: 15}
\begin{gathered}
\frac{\partial^{2} f}{\partial \eta \partial \mu}=0 \Rightarrow \frac{\partial^{2} \mathcal{L}}{\partial \eta \partial \mu}=\frac{\partial^{2} \mathcal{L}}{\partial \eta \partial \dot{\mu}}=\frac{\partial^{2} \mathcal{L}}{\partial \dot{\eta} \partial \mu}=\frac{\partial^{2} \mathcal{L}}{\partial \dot{\eta} \partial \dot{\mu}}=0 \\
\Rightarrow \frac{\delta^{2} \mathcal{A}(x[\tau])}{\delta \eta[t] \delta \mu[t]}=0 \Leftrightarrow(\mu[\tau] \perp \eta[\tau]) \mid b[\tau], x_{0} \\
\frac{\delta \mathcal{A}(x[\tau])}{\delta \mu[t]}=\int d t^{\prime}\left(\frac{\partial \mathcal{L}}{\partial \mu\left[t^{\prime}\right]} \frac{\delta \mu\left[t^{\prime}\right]}{\delta \mu[t]}+\frac{\partial \mathcal{L}}{\partial \dot{\mu}\left[t^{\prime}\right]} \frac{d}{d t^{\prime}} \frac{\delta \mu\left[t^{\prime}\right]}{\delta \mu[t]}+\ldots\right) \\
\frac{\delta^{2} \mathcal{A}(x[\tau])}{\delta \eta[t] \delta \mu[t]}=\int d t^{\prime} d t^{\prime \prime}\left(\frac{\delta \eta\left[t^{\prime \prime}\right]}{\delta \eta[t]} \frac{\partial^{2} \mathcal{L}}{\partial \eta \partial \mu} \frac{\delta \mu\left[t^{\prime}\right]}{\delta \mu[t]}+\frac{\delta \eta\left[t^{\prime \prime}\right]}{\delta \eta[t]} \frac{\partial^{2} \mathcal{L}}{\partial \eta \partial \dot{\mu}} \frac{d}{d t^{\prime}} \frac{\delta \mu\left[t^{\prime}\right]}{\delta \mu[t]}+\ldots\right)
\end{gathered}
\end{equation}
These expressions mean that the probability of an internal path, given a blanket path (and initial states), does not depend on the external path and \textit{vice versa}.

\subsection{Summary}

\begin{figure}
    \centering
    \includegraphics[width=0.7\textwidth]{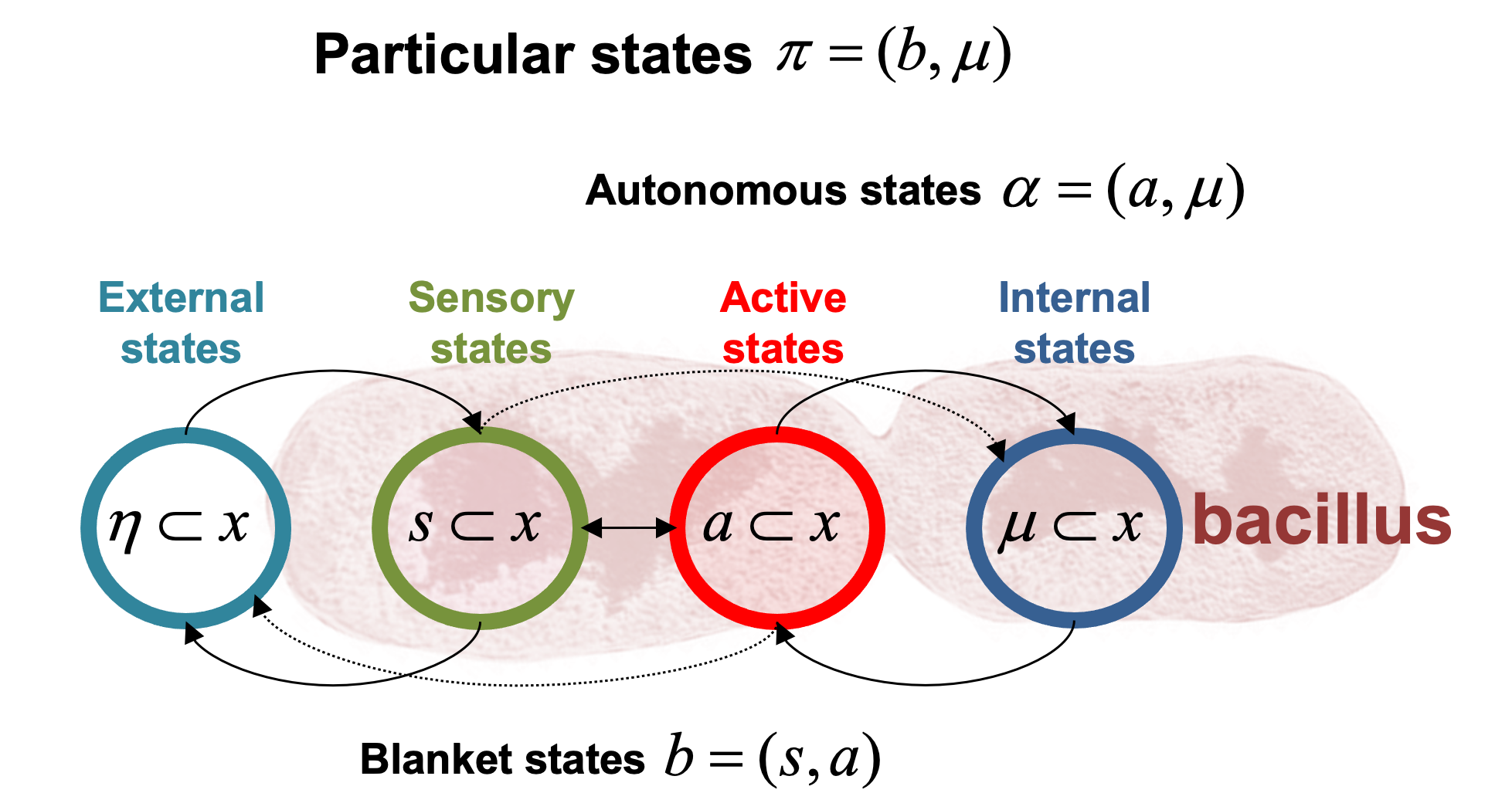}
    \caption{\textbf{Markov blankets}. This influence diagram illustrates a particular partition of states into internal states (blue) and external states (cyan) that are separated by a Markov blanket comprising sensory (green) and active states (red). The edges in this graph represent the influence of one state on another, as opposed to conditional dependencies. The diagram shows this partition as it would be applied to a single-cell organism, where internal states are associated with intracellular states, the sensory states become the surface states or cell membrane overlying active states (e.g., the actin filaments of the cytoskeleton). The dotted lines indicate allowable directed influences from sensory (resp., active) to internal (resp., external) states. Particular states constitute a particle; namely, autonomous and sensory states—or blanket and internal states.}
    \label{fig: 1}
\end{figure}

In summary, the internal dynamics (i.e., paths) of some ‘thing’ are conditionally independent of external paths if, and only if, the flow of internal states does not depend on external states and \textit{vice versa} (given initial states). We take this as a necessary and sufficient condition for something to exist, in the sense that it can be distinguished from everything else. When the initial states are sampled from the NESS density, the internal states are conditionally independent of external states (given blanket states), under certain constraints on solenoidal flow. Figure \ref{fig: 1} illustrates the ensuing particular partition. Note that the edges in this graph represent the influence of one state on another, as opposed to conditional dependencies. This is important because directed influences admit conditional independence. These conditional independencies are manifest as zero entries in the Hessian matrices, which inherit from the sparse, directed coupling of the dynamics.

\section{From self-organisation to self-evidencing}

Equipped with a particular partition, we can now talk about things in terms of their internal states and Markov boundary; namely autonomous states. And we can talk about autonomous states and their Markov boundary; namely, particular states—the states of a particle. The next step is to characterise the flow of the autonomous states (of a particle, plant or person) in relation to external states. In other words, we consider the nature of the coupling between the outside and inside of a particle, across its Markov blanket. It is at this point that we move towards a (Bayesian) mechanics that is the special provenance of systems with particular partitions.

The existence of a particular partition means that—given sensory states—one can stipulatively define the conditional density over external states as being parameterised by the most likely internal state~\cite{dacostaBayesianMechanicsStationary2021a}\footnote{In other words, the internal mode supplies the sufficient statistics of the conditional density over external states.}. We will call this a \textit{variational density} parameterised by the \textit{internal} mode $\boldsymbol{\mu}(\tau)$\footnote{\textbf{Question}: what if the conditional densities are not well-behaved, e.g., what if there are no unique modes? The answer is that well-behaved densities are generally guaranteed when increasing the dimensionality of state-spaces using generalised coordinates of motion \cite{kerrGeneralizedPhaseSpace2000,fristonGeneralisedFiltering2010,fristonPathIntegralsParticular2022}. In other words, instead of just dealing with states, we consider states and their generalised motion to arbitrarily high order. We will see examples of this later.}:
\begin{equation}
\label{eq: 16}
\begin{aligned}
q_{\boldsymbol \mu}(\eta) & \triangleq p(\eta \mid s) \\
\boldsymbol{\alpha}(\tau) &=(\mathbf{a}(\tau), \boldsymbol{\mu}(\tau)) \\
\boldsymbol{\alpha}(\tau) &=\arg \min _{\alpha} \Im(\alpha(\tau) \mid s(\tau)) \Rightarrow \\
\boldsymbol{\alpha}[\tau] &=\arg \min _{\alpha} \mathcal{A}(\alpha[\tau] \mid s[\tau]) \Rightarrow \\
\dot{\boldsymbol{\alpha}}(\tau) &=f_{\alpha}(s, \boldsymbol{\alpha})
\end{aligned}
\end{equation} 						
As with the paths of least action, we will use bold typeface to denote a mode or most likely state, given all the states necessary to specify its likelihood. For autonomous states, we only need the sensory states, because the autonomous states are conditionally independent of external states. 

Inducing the variational density is an important move. It means that for every sensory state there is a corresponding active mode and an internal mode (or an autonomous mode in the joint space of active and internal states). The active $\mathbf{a}(\tau)$, internal $\boldsymbol \mu(\tau)$ and autonomous $\boldsymbol \alpha(\tau)$ modes evolve on \textit{active}, \textit{internal} and \textit{autonomous} \textit{manifolds}\footnote{A manifold is a topological (state-) space where each state has a neighbourhood that is homeomorphic to a portion of an Euclidean space of the same dimension~\cite{leeIntroductionTopologicalManifolds2011}. Intuitively, it is a curved space, such as a smooth surface, in a possibly large but finite number of dimensions. In this instance, the states are conditional modes.}, respectively, whose dimensionality is the same as the sensory states\footnote{The dimensionality of the active, internal and autonomous manifolds corresponds to the number of sensory states. This means that both the number of active and internal states must be greater than the number of sensory states. In turn, this limits the straightforward application of the free energy principle to particular partitions where the number of active states—and the number of internal states—exceeds the number of sensory states. In other words, the FEP applies to large particles with a nontrivial internal dynamics.}. We will see later that these manifolds play the role of \textit{centre manifolds}; namely, manifolds on which dynamics do not diverge (or converge) exponentially fast~\cite{carrApplicationsCentreManifold1982}.

Crucially, the internal manifold is also a \textit{statistical manifold} because its states are sufficient statistics for the variational density. In turn, this means that it is equipped with a metric and implicit information geometry~\cite{parrMarkovBlanketsInformation2020,ayInformationGeometry2017,amariInformationGeometryIts2016}. 
Indeed, the Fisher information metric tensor, which measures changes in the Kullback-Leibler (KL) divergence resulting from infinitesimal changes in the internal mode, is a Riemannian metric that yields an information distance~\cite[Appendix B]{dacostaNeuralDynamicsActive2021}. This means we can interpret dynamics on the internal manifold as updating Bayesian beliefs \textit{about} external states. This interpretation can be unpacked in terms of Bayesian inference as follows.

Equation \eqref{eq: 16} means that for every sensory state there is a conditional density over external states and a corresponding internal mode with the smallest surprisal. This mode specifies the variational density, where—by definition—the KL divergence between the variational density and the conditional density over external states is zero\footnote{Since the variational and conditional densities over external states are equal, any divergence between them will vanish, see \cite[Section 3.2]{amariMethodsInformationGeometry2007}.}. This means we can express the autonomous flow as a gradient flow on a free energy functional of the variational density\footnote{A functional is a function of a function, here, the free energy is a function of a conditional density parameterised by the internal mode.}. From \eqref{eq: 12}
 		\begin{equation}
 		\label{eq: 17}
\left[\begin{array}{l}
f_{\eta}(x) \\
f_{s}(x) \\
f_{a}(\pi) \\
f_{\mu}(\pi)
\end{array}\right]=\Omega\left[\begin{array}{c}
\nabla_{\eta} \Im(x) \\
\nabla_{s} \Im(x) \\
\nabla_{a} F(\pi) \\
\nabla_{\mu} F(\pi)
\end{array}\right]-\Lambda,
\end{equation}							
where the free energy in question is (an upper bound on) the surprisal of particular states:
 \begin{equation}
 \label{eq: 18}
\begin{aligned}
F(\pi(\tau)) &=\mathbb{E}_{q}[\ln q(\eta(\tau))-\ln p(\eta(\tau))-\ln p(\pi(\tau) \mid \eta(\tau))]=\Im(\pi(\tau)) \\
&=\underbrace{\mathbb{E}_{q}[\Im(\eta(\tau), \pi(\tau))]}_{\text {Expected energy }}-\underbrace{\H[q(\eta(\tau))]}_{\text {Entropy }} \\
&=\underbrace{\mathbb{E}_{q}[\Im(\pi(\tau) \mid \eta(\tau))]}_{ \text {-ve Accuracy }}+\underbrace{D[q(\eta(\tau)) \| p(\eta(\tau))]}_{\text {Complexity }} \\
&=\underbrace{D[q(\eta(\tau)) \| p(\eta(\tau) \mid \pi(\tau))]}_{=0}+\Im(\pi(\tau)) \\
q &=q_{\boldsymbol \mu}(\eta)=p(\eta \mid s)=p(\eta \mid \pi) \\
\mathbb{E}[F(\pi)] &=\mathbb{E}[\Im(\pi)]=\H[p(\pi)]
\end{aligned}
\end{equation}				
This variational free energy\footnote{\textbf{Question:} why is this functional called \textit{variational} free energy? More generally (for instance in engineering applications where the free energy in question is also called an evidence lower bound \cite{bishopPatternRecognitionMachine2006}) the free energy is a functional of an approximate posterior density $q$ that is an approximation to the Bayesian posterior, as follows:
\begin{equation}
\label{eq: free energy upper bound}
\begin{split}
q_{\boldsymbol \mu}(\eta)\approx p(\eta \mid \pi) \Rightarrow 
F[q] &=\underbrace{D[q(\eta(\tau)) \| p(\eta(\tau) \mid \pi(\tau))]}_{\geq 0}+\Im(\pi(\tau)) \\
\end{split}
\end{equation}
The variational density considered in this article is the minimiser of \eqref{eq: free energy upper bound}, and the free energy evaluated at the variational density is the variational free energy. The term 'variational' inherits from the use of the calculus of variations in variational Bayes (a.k.a., approximate Bayesian inference), applied in the context of a mean field approximation or factorised form of the variational density. The term 'free energy' inherits from Richard Feynman's path integral formulation, in the setting of quantum electrodynamics.} can be rearranged in several ways. First, it can be expressed as expected \textit{energy} minus the \textit{entropy} of the variational density, which licences the name \textit{free energy}\footnote{\textbf{Question}: is variational free energy the same kind of free energy found in statistical mechanics? The answer is no: the entropy term in the variational free energy is the entropy of a variational density—over external states—parameterised by internal states. This entropy is distinct from the entropy of \textit{internal states}. Minimising variational free energy \textit{increases} the entropy of the variational density and, usually, \textit{reduces} the entropy of internal states (see \cite{ueltzhofferVariationalFreeEnergy2021} for an example). 
Mathematically, we can express the different kind of entropies as $\H[q(\eta(\tau))]\neq \H[p(\mu(\tau))]$.}. In this decomposition, minimising variational free energy corresponds to the maximum entropy principle, under the constraint that the expected energy is minimised~\cite{jaynesInformationTheoryStatistical1957,lasotaChaosFractalsNoise1994}. The expected energy is a functional of the NESS density that plays the role of a \textit{generative model}; namely, a joint distribution over causes (external states) and their consequences (particular states)\footnote{\textbf{Question}: in practical applications, variational free energy is usually a function of data or observed (sensory) states. So, why is variational free energy a function of particular states? Later, we will see that practical applications correspond to Bayesian filtering, under the assumption that particular dynamics are very precise. This means that there is no uncertainty about autonomous paths given sensory paths, and the action of a particular path is the action of a sensory path. In generalised coordinates of motion—used in Bayesian filtering—the action of a path becomes the surprisal of a state. In this setting, the variational free energy of particular states is the same as the variational free energy of sensory states.}.

Second, variational free energy can be decomposed into the (negative) log likelihood of particular states (i.e., negative \textit{accuracy}) and the KL divergence between posterior and prior densities (i.e., \textit{complexity}). Finally, it can be written as the self-information associated with particular states (i.e., \textit{surprisal}) plus the KL divergence between the variational and conditional (i.e., posterior) density, which—by construction—is zero. In variational Bayesian inference~\cite{bealVariationalAlgorithmsApproximate2003}, negative surprisal is read as a log marginal likelihood or model evidence, having marginalised over external states. In this setting, negative free energy is an \textit{evidence lower bound} or ELBO~\cite{winnVariationalMessagePassing2005,bishopPatternRecognitionMachine2006}.

So, in what sense can we interpret \eqref{eq: 17} in terms of inference? Let us start by considering the response of autonomous states to some sensory perturbation: that is, the path of autonomous states conditioned upon sensory states. If sensory states change slowly, then the autonomous states will flow towards their most likely value (i.e., their conditional mode) and stay there\footnote{Or, at least in the vicinity, if there are random fluctuations on its motion.}. However, if sensory states are changing, the autonomous states will look as if they are trying to hit a moving target. One can formulate this along the lines of the centre manifold theorem~\cite{carrApplicationsCentreManifold1982,langVoiceRecognitionAphasic2009}, where we have a (fast) flow \textit{off} the centre manifold and a (slow) flow of the autonomous mode \textit{on} the manifold.
 	\begin{equation}
 	\label{eq: 19}
 \begin{aligned}
\alpha(\tau)&= \underbrace{\varepsilon(\tau)}_{\text{Off manifold}} + \underbrace{\boldsymbol{\alpha}(\tau)}_{\text{On manifold}}\\ \varepsilon(\tau) &\triangleq\alpha(\tau)-\boldsymbol{\alpha}(\tau)
\end{aligned}
\end{equation}	

In effect, this is a decomposition in a frame of reference that moves with the autonomous mode, whose path lies on the centre manifold. 
We further describe the off manifold flow using a Taylor expansion around the (time-varying) autonomous mode\footnote{Note that we are performing a Taylor expansion of a (generally rough) stochastic process $\varepsilon$, see \cite[Chapter 5]{kloedenNumericalSolutionStochastic1992}. Alternatively, it may be possible to instead consider motion in generalised coordinates to introduce smooth random fluctuations (see next Section), so that $\varepsilon$ becomes smooth and the usual Taylor expansion applies.} 

\begin{equation}
\label{eq: 19 bis}
    \begin{aligned}
\dot{\varepsilon}(\tau) &=\dot{\alpha}(\tau)-\dot{\boldsymbol{\alpha}}(\tau)=f_{\varepsilon}(0)+\frac{\partial f_{\varepsilon}}{\partial \varepsilon} \cdot \varepsilon+\ldots=\frac{\partial f_{\alpha}}{\partial \alpha} \cdot \varepsilon+\ldots \\
& \Rightarrow \\
\underbrace{\dot{\alpha}(\tau) -\dot{\boldsymbol{\alpha}}(\tau)}_{\text{Off manifold flow}} &=\mathbf{J}_{\alpha} \cdot(\alpha-\boldsymbol{\alpha})+\ldots \\
&=-\underbrace{\left(\Gamma_{\alpha} \nabla_{\alpha \alpha} F\right) \cdot(\alpha-\boldsymbol{\alpha})}_{\text {Flow to centre manifold }}+\underbrace{\left(Q_{\alpha \alpha} \nabla_{\alpha \alpha} F\right) \cdot(\alpha-\boldsymbol{\alpha})}_{\text {Flow parallel to the manifold}}+\ldots
    \end{aligned}
\end{equation}

\begin{figure}
    \centering
    \includegraphics[width=\textwidth]{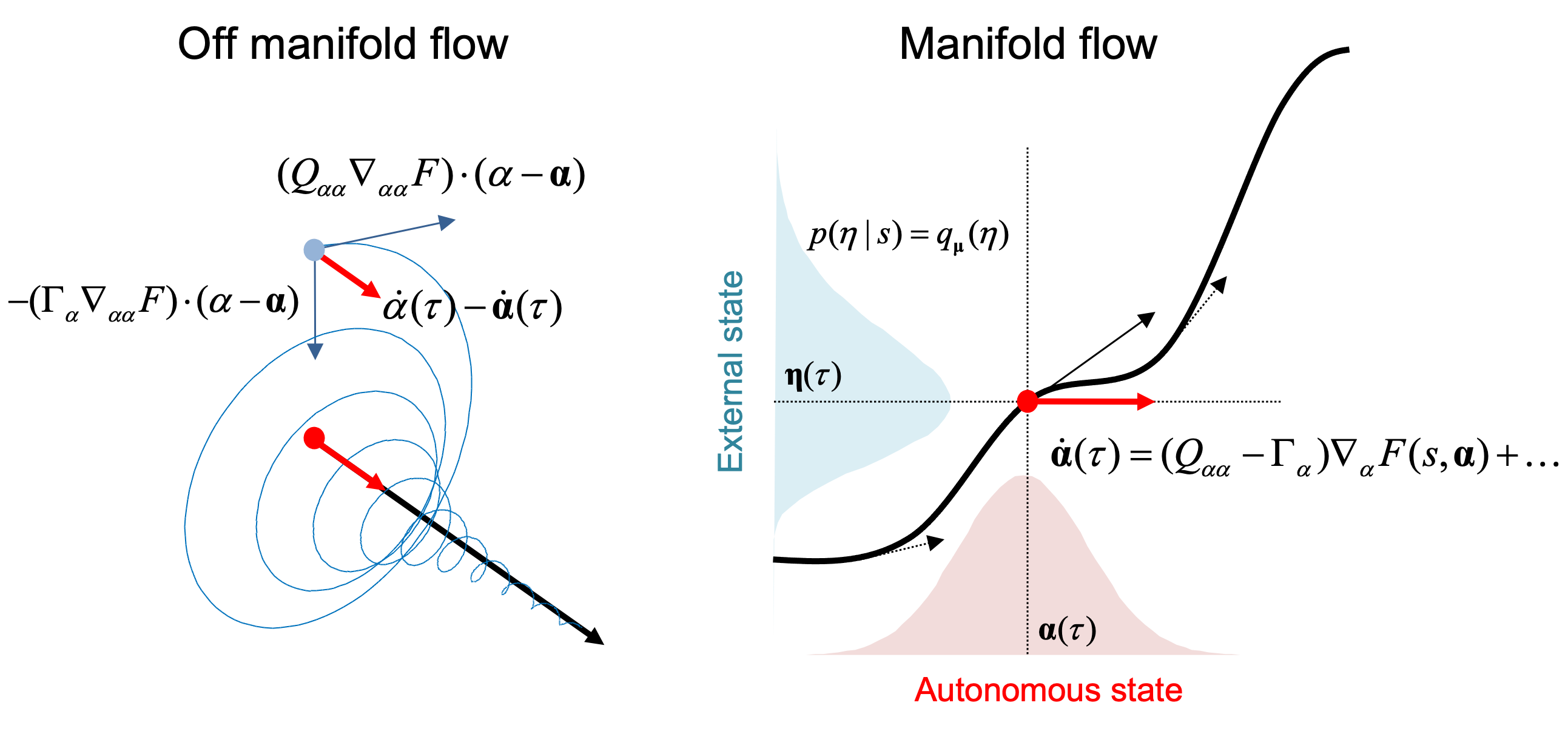}
    \caption{\textbf{Autonomous flows and Bayesian filters}. This figure shows two components of the autonomous flow; namely, a (fast) flow $\dot \alpha(\tau)-\dot {\boldsymbol \alpha}(\tau)$ \textit{off} the (centre) manifold, and a (slow) flow $\dot {\boldsymbol \alpha}(\tau)$ \textit{on} the manifold. The manifold here is the set of autonomous modes $\boldsymbol \alpha(\tau)$ given sensory states $s(\tau)$ for all time $\tau$, see \eqref{eq: 16}. The decomposition into fast and slow flows means that the manifold can be thought of as a centre manifold. The left panel shows two components of the fast flow off the manifold; namely, a flow towards the centre manifold and a flow parallel to the manifold, see \eqref{eq: 19 bis}. This decomposition rests upon a first-order Taylor expansion of the off manifold flow. The right panel plots the external mode as a function of the autonomous mode---what is known as a \textit{synchronisation} manifold---as a black curvilinear line. The Gaussian (blue and red) distributions show possible variations in (external and autonomous) conditional modes due to variations in the sensory states. The arrows represent the centre manifold flow $\dot {\boldsymbol \alpha}(\tau)$ in the context of this synchronisation manifold, where the tangent vectors represent possible directions of the flow.
    }
    \label{fig: 2}
\end{figure}

This means that the flow at the expansion point is zero, leaving the second term of the expansion as the first non-vanishing term. This is the Jacobian of the autonomous flow times the displacement of the current autonomous state from its corresponding mode. The second-order derivatives of the free energy arise from the Jacobian of the flow, i.e., substituting \eqref{eq: 17} into \eqref{eq: 8}. Therefore, the off manifold flow has a component that flows \textit{towards} the centre manifold,\footnote{We know that the flow must be towards the centre manifold because the covariance of random fluctuations is positive definite, and the curvature of the free energy is positive definite at its minima: i.e., around the expansion point.} afforded by the gradient flow, and a component that is \textit{parallel} to the manifold, afforded by the solenoidal flow, cf. \eqref{eq: 6}. Taken together, this means that the autonomous states flow in ever-decreasing circles towards the centre manifold, as illustrated in Figure \ref{fig: 2}.

But what about the flow \textit{on} the centre manifold? We know from \eqref{eq: 17} that the flow of the autonomous mode can be expressed in terms of free energy gradients:
\begin{equation}
\label{eq: 20}
\begin{aligned}
\dot{\boldsymbol{\alpha}}(\tau) &=\left(Q_{\alpha \alpha}-\Gamma_{\alpha}\right) \nabla_{\alpha} F(s, \boldsymbol{\alpha})+\ldots \\
&=\left(Q_{\alpha \alpha}-\Gamma_{\alpha}\right) \underbrace{\nabla_{\alpha} \mathbb{E}_{q}[\Im(s, \boldsymbol{\alpha} \mid \eta)]}_{\text {-ve Accuracy }}+\left(Q_{\alpha \alpha}-\Gamma_{\alpha}\right) \underbrace{\nabla_{\alpha} D\left[q_{\boldsymbol \mu}(\eta) \| p(\eta)\right]}_{\text {Complexity }}+\ldots
\end{aligned}
\end{equation}
 			
This expression unpacks the centre manifold flow in terms of the accuracy and complexity parts of free energy, where the accuracy part depends upon the sensory states, while the complexity part is a function of, and only of, autonomous states. In short, the flow on the centre manifold will look as if it is trying to maximise the accuracy of its predictions, while complying with prior (Bayesian) beliefs.\footnote{The covariance of random fluctuations $\Gamma_\alpha$ is positive definite and the solenoidal matrix field $Q_{\alpha\alpha}$ is skew-symmetric, therefore the flow in \eqref{eq: 20} will seek to minimise complexity minus accuracy.} Here, predictions are read as the expected sensory states, under posterior (Bayesian) beliefs about their causes afforded by the variational density over external states. 

\subsection{Summary}
In summary, a particular partition of a nonequilibrium steady-state density implies that autonomous dynamics can be interpreted as performing a particular kind of inference. This entails a fast flow towards an autonomous centre manifold and a slow flow on the centre manifold. The centre manifold flow can be interpreted as Bayesian belief updating, where posterior (Bayesian) beliefs are encoded by points on an internal (statistical) manifold. In other words, for every point on the statistical manifold, there is a corresponding variational density or Bayesian belief over external states. We are now in a position to express this belief updating as a variational principle of least action:

\begin{equation}
\label{eq: 21}
\begin{aligned}
\boldsymbol{\alpha}[\tau] &=\arg \min _{\alpha[\tau]} \mathcal{A}(\alpha[\tau] \mid s[\tau]) \\
& \Leftrightarrow \delta_{\alpha} \mathcal{A}(\boldsymbol{\alpha}[\tau] \mid s[\tau])=0 \\
& \Leftrightarrow \\
\dot{\mathbf{a}}(\tau) &=f_{a}(s, \boldsymbol{\alpha})=\left(Q_{a a}-\Gamma_{a}\right) \nabla_{a} F(s, \boldsymbol{\alpha})+\ldots \\
\dot{\boldsymbol{\mu}}(\tau) &=f_{\mu}(s, \boldsymbol{\alpha})=\left(Q_{\mu \mu}-\Gamma_{\mu}\right) \nabla_{\mu} F(s, \boldsymbol{\alpha})+\ldots
\end{aligned}
\end{equation}
 							
This is a basis of the free energy principle. Put simply, it means that the internal states of a particular partition can be cast as encoding conditional or posterior Bayesian beliefs \textit{about} external states. Equivalently, the autonomous path of least action can be expressed as a gradient flow on a variational free energy that can be read as log evidence. This licences a somewhat poetic description of self-organisation as self-evidencing~\cite{hohwySelfEvidencingBrain2016}, in the sense that the surprisal or self-information is known as log model evidence or marginal likelihood in Bayesian statistics\footnote{\textbf{Question}: this Bayesian mechanics seems apt for inference but what about learning over time? We have been dealing with states in a generic sense. However, one can have states that change over different timescales. One can read slowly changing states as special states that play the role of parameters; either parameters of the flow or, implicitly, the generative model. In mathematical and numerical analyses, states and parameters are usually treated identically; i.e., as minimising variational free energy. Indeed, in practical applications of Bayesian filtering schemes that learn, the parameters are treated as slowly changing states. See \cite{schiffKalmanFilterControl2008,fristonGeneralisedFiltering2010} for worked examples.}.

Interestingly, because of the symmetric setup of the Markov blanket, it would be possible to repeat everything above but switch the labels of internal and external states—and active and sensory states—and tell the same story about external states tracking internal states. This evinces a form of generalised synchrony~\cite{huntDifferentiableGeneralizedSynchronization1997,jafriGeneralizedSynchronyCoupled2016,buendiaBroadEdgeSynchronization2022,dacostaBayesianMechanicsStationary2021a}, where internal and external states track each other. Technically, if we consider the (internal and external) manifolds in the joint space of internal and external states, we have something called a synchronisation manifold that offers another perspective on the coupling between the inside and outside~\cite{fristonActiveInferenceCommunication2015,parrMarkovBlanketsInformation2020,dacostaBayesianMechanicsStationary2021a}. 

\begin{figure}
    \centering
    \includegraphics[width=0.8\textwidth]{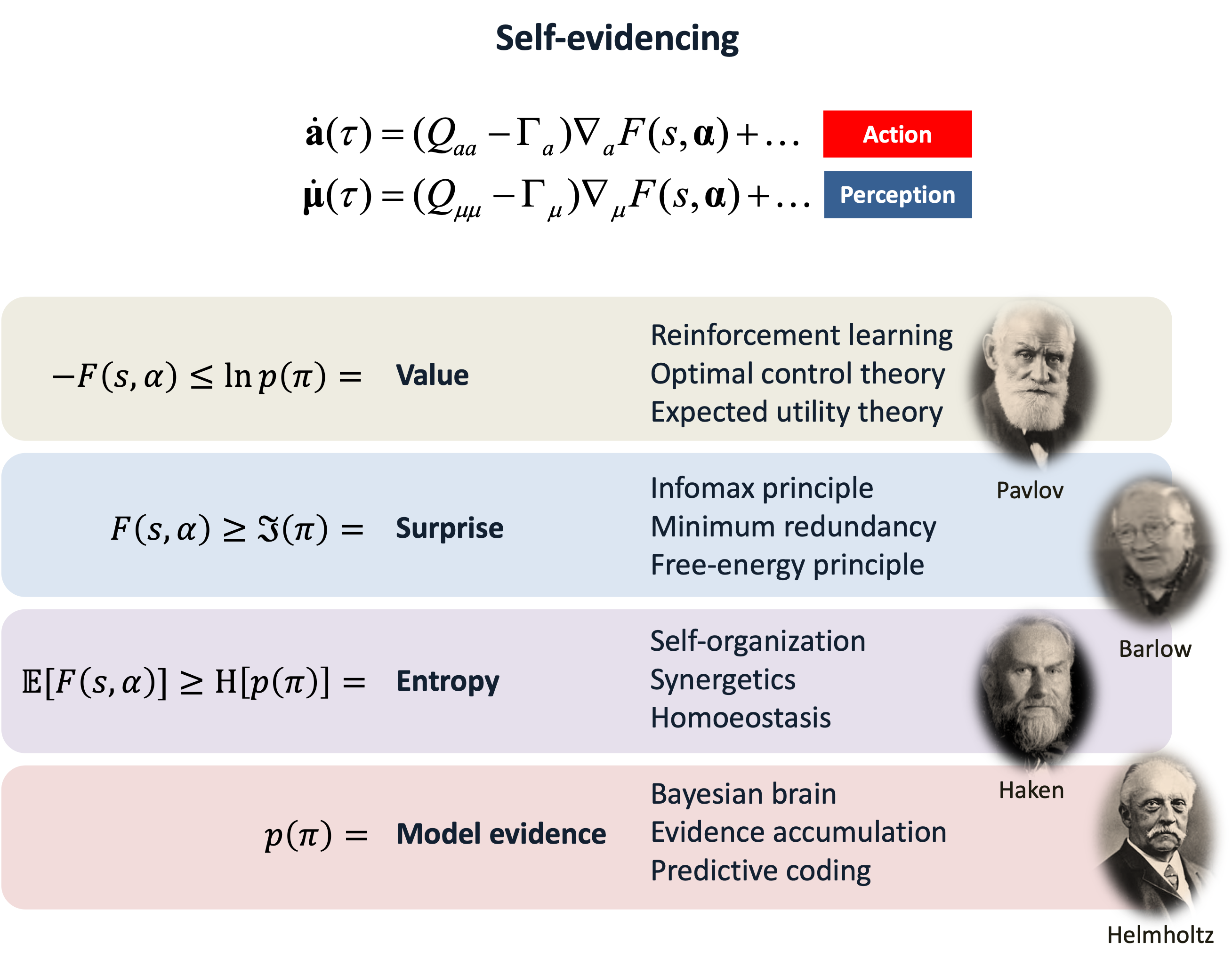}
    \caption{\textbf{Markov blankets and self-evidencing}. This schematic illustrates various points of contact between minimising variational free energy and other normative theories of optimal behaviour. The existence of a Markov blanket entails a certain lack of influences among internal, blanket and external states. These have an important consequence—internal and active states are not influenced by external states, which means their dynamics (i.e., perception and action) are a function of, and only of, particular states, given by a variational (free energy) bound on surprisal. This has a number of interesting interpretations. Given surprisal is the negative log probability of finding a particle or creature in a particular state, minimising surprise corresponds to maximising the value of that state. This interpretation is licensed by the fact that the states with a high probability are, by definition, characteristic of the particle in question. On this view, one could relate this to dynamics in reinforcement learning~\cite{bartoReinforcementLearningIntroduction1992}, optimal control theory~\cite{todorovOptimalFeedbackControl2002} and, in economics, expected utility theory~\cite{bossaertsBehaviouralEconomicsNeuroeconomics2015a,vonneumannTheoryGamesEconomic1944}. Gradient flows that minimise surprisal (i.e., self-information) lead to a series of influential accounts of neuronal dynamics; including the principle of maximum mutual information~\cite{opticanTemporalEncodingTwodimensional1987a,linskerPerceptualNeuralOrganization1990}, the principles of minimum redundancy and maximum efficiency~\cite{barlowPossiblePrinciplesUnderlying1961} and the free energy principle~\cite{fristonFreeEnergyPrinciple2006}. Crucially, the average or expected surprise (over time of particular states) corresponds to entropy. This means that action and perception look as if they are bounding the entropy of particular states. This links nicely with theories of self-organisation, such as synergetics in physics~\cite{nicolisSelforganizationNonequilibriumSystems1977,hakenSynergeticsIntroductionNonequilibrium1978,kauffmanOriginsOrderSelforganization1993} or homoeostasis in physiology~\cite{ashbyPrinciplesSelfOrganizingDynamic1947,conantEveryGoodRegulator1970,bernardLecturesPhenomenaLife1974}. Finally, the probability of a particular state, is, on a statistical view, model evidence or marginal likelihood~\cite{mackayFreeEnergyMinimisation1995,mackayInformationTheoryInference2003}, marginalising over the causes of particular states (i.e., external states). This means that all the above formulations are internally consistent with things like the Bayesian brain hypothesis, evidence accumulation and predictive coding\cite{fristonFreeenergyPrincipleUnified2010,bogaczTutorialFreeenergyFramework2017,buckleyFreeEnergyPrinciple2017}. Most of these formulations inherit from Helmholtz's motion of unconscious inference~\cite{helmholtzHelmholtzTreatisePhysiological1962}, later unpacked in terms of perception as hypothesis testing in psychology~\cite{gregoryPerceptionsHypotheses1980} and machine learning~\cite{dayanHelmholtzMachine1995}. Although not depicted here, the minimisation of complexity—inherent in the minimisation of free energy—enables thermodynamic and metabolic efficiency via Landauer's principle~\cite{landauerIrreversibilityHeatGeneration1961}.}
    \label{fig: 3}
\end{figure}

These teleological interpretations cast particular paths of least action as an optimisation process, where different readings of free energy link nicely to various normative (i.e., optimisation) theories of sentient behaviour. Some cardinal examples are summarised in Figure \ref{fig: 3}; see \cite{fristonFreeenergyPrincipleUnified2010,buckleyFreeEnergyPrinciple2017,fristonFreeEnergyPrinciple2019a,fristonReinforcementLearningActive2009} for some formal accounts of these relationships. Because internal states do not influence sensory (or external) states, they will look as if they are concerned purely with inference, in the sense that they parameterise the variational density over external states. However, active states influence sensory (and external) states and will look as if they play an active role in configuring (and causing) the sensory states that underwrite inference. In the neurosciences, this is known as \textit{active inference}~\cite{fristonActionBehaviorFreeenergy2010,ueltzhofferDeepActiveInference2018,koudahlWorkedExampleFokkerPlanckBased2020}.

The link between optimisation and inference is simply that inference is belief optimisation. However, it is worth unpacking the gradients that ‘drive’ this optimisation. In statistics, variational free energy is used to score the divergence between a variational density and the conditional density over external (i.e., hidden) states, given blanket states~\cite{winnVariationalMessagePassing2005}. Unlike the definition in \eqref{eq: 18}, these densities are not assumed to be equivalent. Variational inference proceeds by optimising the variational density such that it minimises free energy—often using the gradient flows in \eqref{eq: 17}. However, there is a subtle difference between the dynamics of \eqref{eq: 17} and variational inference. In the former, there is no contribution from the KL-divergence as it is stipulated to be zero. In the latter, it is only the divergence term that contributes to free energy gradients. So, is it tenable to interpret gradient flows on variational free energy as variational inference, or is this just teleological window-dressing? The next section addresses this question through the lens of Bayesian filtering. In brief, we will see that the autonomous paths of least action—implied by a particular partition—are the paths of least action of a Bayesian filter. This takes us beyond ‘as if’ arguments by establishing a formal connection between particular dynamics and variational inference.

\section{Lagrangians, generalised states and Bayesian filtering}

Now, say we wanted to emulate or simulate active inference. Given some equations of motion and statistics of random fluctuations, we could find the stationary solution to the Fokker Planck equation and accompanying Helmholtz decomposition. We could then solve \eqref{eq: 21} for the autonomous paths of least action that characterise the expected behaviour of this kind of particle, and obtain realisations of synchronisation and inference. See \cite{fristonStochasticChaosMarkov2021} for a worked example using a system of coupled Lorentz attractors.


In this section, we take a somewhat pragmatic excursion to suggest a simpler way to recover the paths of least action; namely, as the solution to a generic (Bayesian) filtering scheme that is widely used in the engineering literature.

\subsection{Dynamics in generalised coordinates of motion}

Let us go back to the Langevin equation governing our system 

\begin{equation}
\label{eq: langevin fluctuating covariance}
    \dot x(\tau) = f(x) + \omega(\tau).
\end{equation}
In this section, we assume that the random fluctuations driving the motion have smooth (analytic) sample paths; thus, the Langevin equation considered in the rest of the article can be seen as the limit of \eqref{eq: langevin fluctuating covariance} as the fluctuations become rough \cite{wongRelationOrdinaryStochastic1965}. This setup speaks nicely to the fact that, in biology, fluctuations are often smooth up to a certain order---contrariwise to thermal (white noise) fluctuations---as they are the output of other random dynamical systems. 
As before, we assume that the fluctuations are state-independent, and a stationary Gaussian process, e.g., the smoothing of white noise fluctuations with a Gaussian kernel. Just like in the case of white noise, Gaussianity can be motivated by the central limit theorem---fluctuations should be normally distributed at each point in time.

We denote the autocovariance of fluctuations by 
$\Gamma_{h}= \frac 1 2\mathbb E[\omega (\tau) \otimes \omega (\tau+h)].$
The underlying dynamical systems giving rise to this generic type of smooth noise can be recovered through a procedure known as stochastic realisation \cite{dacostaBayesianMechanicsStationary2021a,lindquistRealizationTheoryMultivariate1985,mitterTheoryNonlinearStochastic1981}. The solution to the Langevin equation 
\eqref{eq: langevin fluctuating covariance} can be approximated, on a suitably small interval of time, by a linear Langevin equation in generalised coordinates of motion $\vec{x}=\left(x, x^{\prime}, x^{\prime \prime}, \ldots\right)$ \cite[Section 4]{balajiBayesianStateEstimation2011}:\footnote{\label{footnote expansion in gen coords}The expansion \eqref{eq: 22} is a linear approximation of \eqref{eq: langevin fluctuating covariance} \cite{biscayLocalLinearizationMethod1996}, obtained by recursively differentiating \eqref{eq: langevin fluctuating covariance} and ignoring the contribution of the derivatives of the flow of order higher than one. In other words, the expansion is exact when the flow is linear, and it is accurate on a short time-scale when the flow is non-linear.}\footnote{The curvature (i.e., second derivative) of the autocovariance $\Gamma^{\prime\prime}_{0}$ is a ubiquitous measure of roughness of a stochastic process \cite{coxTheoryStochasticProcesses1977}. Note that in the limit where the fluctuations $\omega$ are uncorrelated (e.g., white noise fluctuations), $\Gamma^{\prime\prime}_{0}$ (and higher derivatives) become infinitely large.}

\begin{equation}
 \label{eq: 22}
 \begin{split}
     \left.\begin{array}{c}
\dot{x}=x^{\prime}=\nabla f \cdot x+ \omega \\
\dot{x}^{\prime}=x^{\prime \prime}=\nabla f \cdot x^{\prime}+\omega^{\prime} \\
\dot{x}^{\prime \prime}=x^{\prime \prime \prime}=\nabla f \cdot x^{\prime \prime}+\omega^{\prime \prime} \\
\vdots
\end{array}\right\} &\Leftrightarrow \begin{array}{r}
\dot{\vec{x}}=\bold f (\vec{x})+\vec{\omega} \\
\mathbf{D} \vec{x}
=\mathbf{J} \vec{x}+\vec{\omega} \\
p(\vec{\omega}(\tau))=\mathcal{N}(\vec{\omega}(\tau); 0,2 \boldsymbol{\Gamma})
\end{array}\\
\mathbf{D}=\left[\begin{array}{llll}
0&1 & & \\
&0& 1 & \\
& & 0& \ddots \\
& & & \ddots
\end{array}\right], \quad \mathbf{J}=&\left[\begin{array}{llll}
\nabla f & & & \\
& \nabla f & & \\
& & \nabla f & \\
& & & \ddots
\end{array}\right], \quad \boldsymbol \Gamma=\left[\begin{array}{cccc}
\Gamma_{0} & & \Gamma^{\prime\prime}_{0} & \\
& -\Gamma^{\prime\prime}_{0} & & -\Gamma^{\prime \prime \prime\prime}_{0} \\
\Gamma^{\prime\prime}_{0} & & \Gamma^{\prime \prime \prime\prime}_{0} & \\
& -\Gamma^{\prime \prime \prime\prime}_{0} & & \ddots
\end{array}\right]
 \end{split}
\end{equation}


Here, the different variables $\vec x=\left(x, x^{\prime}, x^{\prime \prime}, \ldots, x^{(n)}, \ldots \right)$ can be seen as the position, velocity, acceleration, jerk, and higher orders of motion of the process, which are treated as separate (generalised) states that are coupled through the Jacobian $\mathbf{J}$. These are driven by smooth fluctuations $\vec{\omega}$ (i.e., the serial derivatives of $\omega$) whose covariance $2\boldsymbol \Gamma$ can be expressed in terms of the serial derivatives of the autocovariance \cite[Appendix A.5.3]{parrActiveInferenceFree2022}.

The generalised states are the coefficients of a Taylor series expansion of the solution to the Langevin equation \eqref{eq: langevin fluctuating covariance}:

\begin{equation}
\label{eq: analytic solution}
   x(\tau) = x(0) + x^{\prime}(0)\tau + \frac {x^{\prime \prime}(0)}2 \tau^2 + \ldots + \frac{x^{(n)}(0)}{n!}\tau^n +\ldots,
\end{equation}
where \eqref{eq: analytic solution} holds, typically, only on a small time-interval to which we restrict ourselves henceforth. In other words, the generalised states at any time-point determine the system's trajectory, and vice versa; that is, there is an isomorphism between generalised states and paths.




 

This line of reasoning has two advantages. First, it means one can let go of white noise assumptions on the random fluctuations and deal with smooth or analytic fluctuations. Second, the linear expansion in generalised coordinates of motion \eqref{eq: 22} means that the distribution of generalised states has a simple Gaussian form
 \begin{equation}
 \label{eq: 23}
\begin{aligned}
\mathcal{L}(\vec{x}(\tau)) &\triangleq -\ln p (\vec{x}(\tau))=\frac{1}{2} \vec{x}(\tau) \cdot \mathbf{M} \vec{x}(\tau) \\
\mathbf{M} &=(\mathbf{D}-\mathbf{J}) \cdot \frac{1}{2 \boldsymbol \Gamma}(\mathbf{D}-\mathbf{J}).
\end{aligned}
\end{equation}
Here, $\mathbf{M}$ can be read as a mass matrix. This suggests that precise particles, with low amplitude random fluctuations, behave like massive bodies. Furthermore, \eqref{eq: 23} is seen as the Lagrangian in generalised coordinates of motion, due to its formal similarity with \eqref{eq: action}.
Under the isomorphism between points and paths in generalised coordinates, the Lagrangian is equivalent to the action; it scores the likelihood of paths of \eqref{eq: langevin fluctuating covariance}, as a path corresponds to a point in generalised coordinates of motion\footnote{\textbf{Question}: how can a point be a path? The generalised states (i.e., temporal derivatives) approximate the path of the solution to \eqref{eq: langevin fluctuating covariance} on a suitably small time interval because they are the coefficients of a Taylor expansion of the path as a function of time \eqref{eq: analytic solution}.}. We will reason about the trajectories of the system by analysing the Lagrangian of generalised states henceforth.

The path of least action corresponds to the minimiser of the Lagrangian, which can be expressed as follows:
 \begin{equation}
  \label{eq: 25}
\begin{aligned}
\overrightarrow{\mathbf{x}}(\tau) &=\arg \min _{\vec{x}(\tau)} \mathcal{L}(\vec{x}(\tau)) \\
& \Leftrightarrow \nabla \mathcal{L}(\overrightarrow{\mathbf{x}}(\tau))=0  \quad \forall \tau .
\end{aligned}
\end{equation}									

We can recover the path of least action by solving the following equation of motion
 \begin{equation}
 \label{eq: 24}
\begin{aligned}
\dot{\vec{x}}(\tau) &=\mathbf{D} \vec{x}-\nabla \mathcal{L}(\vec{x}) \\
\nabla \cdot \mathbf{D} \vec{x} &=0.
\end{aligned}
\end{equation}	
Indeed, this motion can be interpreted as a gradient descent on the Lagrangian, in a frame of reference that moves with the mode of the distribution of generalised states~\cite{fristonGeneralisedFiltering2010}. Thus, the convexity of the Lagrangian means that any solution to \eqref{eq: 24} converges to the path of least action.
In this setting, the divergence-free flow (i.e., the first term) is known as a \textit{prediction} of the generalised state based upon generalised motion, while the curl-free, gradient flow (i.e., the second term) is called an \textit{update}.

\subsection{Particular partitions in generalised coordinates of motion}

We now reintroduce the distinction between internal, external, sensory and active states $x=(\eta, s, a , \mu)$. Briefly, as before, we assume that the Langevin equation \eqref{eq: langevin fluctuating covariance} is sparsely coupled as in \eqref{eq: 13}. This implies that the trajectories internal and external to the particle are conditionally independent given the trajectories of the blanket \eqref{eq: 15}. The same sparse coupling structure carries through the expansion in generalised coordinates \eqref{eq: 22} so that the motion of generalised states entails trajectories with the same conditional independencies. Since paths correspond to generalised states, this yields conditional independence between generalised states, as follows:
\begin{equation}
    (\vec \mu \perp \vec\eta) \mid \vec b \iff \mathcal{L}(\vec{x}) = \mathcal{L}(\vec{\eta} \mid \vec{b}) + \mathcal{L}(\vec{\mu} \mid \vec{b}) + \mathcal{L}(\vec{b}).
\end{equation}

We can now recover paths of least action of the particle by equating the Lagrangian with the variational free energy of generalised states. This allows us to express the internal path of least action as a gradient flow on variational free energy, which can itself be expressed in terms of generalised prediction errors. 
From \eqref{eq: 24}, we have
\begin{equation}
 \label{eq: 26}
\begin{aligned}
\dot{\vec{\mu}}(\tau) & =\mathbf{D} \vec{\mu}-\nabla_{\vec{\mu}} \mathcal{L}(\vec{x})=\mathbf{D} \vec{\mu}-\nabla_{\vec{\mu}} \mathcal{L}(\vec{\mu} \mid \vec{b}) \\
& =\mathbf{D} \vec{\mu}-\nabla_{\vec{\mu}} \mathcal{L}(\vec{\pi}) =\mathbf{D} \vec{\mu}-\nabla_{\vec{\mu}} F(\vec{\pi}),
\end{aligned}
\end{equation}
where the free energy of generalised states is analogous to \eqref{eq: 18}
\begin{equation}
\label{eq: free energy generalised}
    \begin{aligned}
        F(\vec{s}, \vec{a}, \vec{\mu}) & =\underbrace{\mathbb{E}_q[\mathcal{L}(\vec{\eta}, \vec{\pi})]}_{\text {Expected energy }}-\underbrace{H[q(\vec{\eta})]}_{\text {Entropy }} \\
& =\underbrace{\mathbb{E}_q[\mathcal{L}(\vec{\pi} \mid \vec{\eta})]}_{\text {Accuracy }}+\underbrace{D[q(\vec{\eta}) \| p(\vec{\eta})]}_{\text {Complexity }} \\
& =D[q(\vec{\eta}) \| p(\vec{\eta} \mid \vec{\pi})]+\mathcal{L}(\vec{\pi})\\
q_{\vec{\mu}}(\vec{\eta}) & =\mathcal{N}(\vec{\eta}; \overrightarrow{\boldsymbol{\mu}}, \Sigma(\overrightarrow{\boldsymbol{\mu}}))=p(\vec{\eta} \mid \vec{\pi})=p(\vec{\eta} \mid \vec{b}) \\
\mathcal{L}(\vec{\eta}, \vec{\pi}) & =\varepsilon_{\vec{\eta}} \cdot \frac{1}{4 \boldsymbol \Gamma_\eta} \varepsilon_{\vec{\eta}}+\varepsilon_{\vec{s}} \cdot \frac{1}{4 \boldsymbol \Gamma_s} \varepsilon_{\vec{s}}+\ldots \\
\varepsilon_{\vec{\eta}} & \triangleq\mathbf{D} \vec{\eta}-\bold f_{\vec{\eta}}(\vec{\eta}, \vec{s}) \\
\varepsilon_{\vec{s}} & \triangleq\mathbf{D} \vec{s}-\bold f_{\vec{s}}(\vec{\eta}, \vec{s}).
    \end{aligned}
\end{equation}
The variational free energy of generalised states is easy to evaluate, given a generative model in the form of a state-space model~\cite{fristonGeneralisedFiltering2010}; that is, the generalised flow of external and sensory states $\bold f_{\vec{\eta}},\bold f_{\vec{s}}$, and the covariance of their generalised fluctuations 
$\boldsymbol \Gamma_\eta, \boldsymbol \Gamma_s$. Note that the parameterisation of the variational density is very simple: the internal states parameterise the expected external states. Furthermore, the quadratic form of the Lagrangian means that the variational density over the generalised motion of external states is Gaussian\footnote{\textbf{Question}: why is the covariance of the variational density only a function of the internal mode? This follows from the quadratic Lagrangian that furnishes an analytic solution to the free energy minimum. Please see \cite{fristonVariationalFreeEnergy2007} for details.}. This licenses a ubiquitous assumption in variational Bayes called the Laplace assumption. Please see~\cite{fristonVariationalFreeEnergy2007} for a discussion of the simplifications afforded by the Laplace assumption. 

Crucially, in the absence of active states, the dynamic in \eqref{eq: 26} coincides with a generalised Bayesian filter. Generalised filtering is a generic Bayesian filtering scheme for nonlinear state-space models formulated in generalised coordinates of motion~\cite{fristonGeneralisedFiltering2010}; 
special cases include variational filtering \cite{fristonVariationalFiltering2008a}, dynamic expectation maximisation \cite{fristonVariationalTreatmentDynamic2008}, extended Kalman filtering \cite{loeligerLeastSquaresKalman2002}, and generalised predictive coding. 

Furthermore, if the autonomous paths are conditionally independent from external paths, given sensory paths\footnote{This is the case for precise particles, which are defined by particular fluctuations of infinitesimally small amplitude---see next Section and \cite{fristonPathIntegralsParticular2022}.}, the autonomous paths of least action can be recovered from a generalised gradient descent on variational free energy:
\begin{equation}
 \label{eq: gen grad descent autonomous states}
\begin{aligned}
\dot{\vec{\alpha}}(\tau) & =\mathbf{D} \vec{\alpha}-\nabla_{\vec{\alpha}} \mathcal{L}(\vec{x})=\mathbf{D} \vec{\alpha}-\nabla_{\vec{\alpha}} \mathcal{L}(\vec{\alpha} \mid \vec{s}) \\
& =\mathbf{D} \vec{\alpha}-\nabla_{\vec{\alpha}} \mathcal{L}(\vec{\pi}) =\mathbf{D} \vec{\alpha}-\nabla_{\vec{\alpha}} F(\vec{\pi}).
\end{aligned}
\end{equation}
In this case, the most likely paths of both internal and active states can be recovered by a gradient descent on variational free energy, and one can simulate active inference using generalisations of linear quadratic control or model predictive control \cite{kappenPathIntegralsSymmetry2005,todorovGeneralDualityOptimal2008a}:

\begin{equation}
\label{eq: 27}
\left[\begin{array}{c}
\dot{\vec{\eta}}(\tau) \\
\dot{\vec{s}}(\tau) \\
\dot{\vec{ a}}(\tau) \\
\dot{\vec{ \mu}}(\tau)
\end{array}\right]=\left[\begin{array}{c}
\bold f_{\vec{\eta}}(\vec{\eta}, \vec{s}, \vec{a})+\vec{\omega}_{\eta}(\tau) \\
\bold f_{\vec{s}}(\vec{\eta}, \vec{s}, \vec{a})+\vec{\omega}_{s}(\tau) \\
\mathbf{D} \vec{ a}-\nabla_{\vec{a}} F(\vec{s}, \vec{a}, \vec{\mu}) \\
\mathbf{D} \vec{\mu}-\nabla_{\vec{\mu}} F(\vec{s}, \vec{  a}, \vec{ \mu})
\end{array}\right]
\end{equation}
 								
This is effectively a (generalised) version of the particular dynamics in \eqref{eq: 21}.

\subsection{Summary}

This section has taken a somewhat pragmatic excursion from the FEP narrative to consider generalised coordinates of motion. This excursion is important because it suggests that the gradient flows in systems with attracting sets are the paths of least action in Bayesian filters used to assimilate data in statistics~\cite{loeligerLeastSquaresKalman2002} and, indeed, control theory~\cite{vandenbroekRiskSensitivePath2010}. 

Working in generalised coordinates of motion is effectively working with paths and the path integral formulation. Practically, this is useful because one can use the density over paths directly to evaluate the requisite free energy gradients, as opposed to solving the Fokker-Planck equation to find the NESS density. Effectively, the generative model becomes a state-space model, specified with flows and the statistics of random fluctuations: see \eqref{eq: free energy generalised}. These are the sufficient statistics of the joint density over external and sensory paths.

Hitherto, we have largely ignored random fluctuations in the motion of particular states to focus on the underlying flows. Are these flows ever realised or does the principle of least action in \eqref{eq: 21} only apply to the most likely autonomous paths? In what follows, we will consider a special class of systems, where we suppress particular fluctuations to recover the behaviour of particles that show a precise or predictable response to external states. For this kind of particle, the particular paths are always the paths of least action.

\section{From statistical to classical particles}
\label{sec: precise}

So far, we have a Bayesian mechanics that would be apt to describe a particle or person with a pullback attractor. But what is the difference between a particle and a person? This question speaks to distinct classes of things to which the free energy principle could apply; e.g., molecular versus biological. Here, we associate biotic self-organisation with the precise and predictable dynamics of large particles. Thanks to the Helmholtz decomposition \eqref{eq: 6}, it is known that when random fluctuations are large, dissipative flow dominates conservative flow, and we have ensembles described by statistical mechanics (i.e., small particles). Conversely, when random fluctuations have a low amplitude, solenoidal flow\footnote{And its accompanying correction term $\Lambda$, see \eqref{eq: 6}.} dominates and we have classical mechanics and deterministic chaos (i.e., of heavenly and $n$-body problems). Here, we consider the distinction between statistical and classical mechanics in the setting of a particular partition.

It is often said that the free energy principle explains why biological systems resist the second law and a natural tendency to dissipation and disorder~\cite{fristonLifeWeKnow2013}. However, this is disingenuous on two counts. First, the second law only applies to closed systems, while the free energy principle describes open systems in which internal states are exposed to—and exchange with—external states through blanket states. Second, there is nothing, so far, to suggest that the entropy of particular states or paths is small. Everything we have done would apply equally to particles with high and low entropy densities. So, what distinguishes between high and low entropy systems (e.g., between candle flames and concierges), respectively? 

One answer can be found in the path-integral formulation: from \eqref{eq: 5}, we can associate the entropy of a path (i.e., history or trajectory of particular states) with the amplitude of random fluctuations. This licences the notion of \textit{precise particles} that are characterised by low or vanishing random fluctuations\footnote{\textbf{Question}: but surely my neurons are noisy? There is a substantial literature that refers to neuronal and synaptic noise: e.g., \cite{toutounjiSpatiotemporalComputationsExcitable2014}. However, the population dynamics of neuronal ensembles or assemblies are virtually noiseless by the central limit theorem (because they comprise thousands of neurons), when averaged over suitable spatial and temporal scales. For example, in electrophysiology, averaging several fluctuating single trial responses yields surprisingly stable and reproducible event-related potentials. From the perspective of the FEP, studying single neurons (or trials) is like studying single molecules to characterise fluid dynamics.}. In essence, precise particles are simply ‘things’ that are subject to the classical laws of nature; i.e., Lagrangian mechanics. In the accompanying limit of small fluctuations on particular states, every autonomous trajectory is a path of least action. From \eqref{eq: 5} and \eqref{eq: 21} this can be expressed as follows:
 \begin{equation}
 \label{eq: 28}
\begin{aligned}
\lim _{\Gamma_{\pi} \rightarrow 0} \dot{\alpha}(\tau) &=f_{\alpha}(\pi(\tau)) \Leftrightarrow \delta_{\alpha} \mathcal{A}(\alpha[\tau] \mid s[\tau])=0 \Leftrightarrow \alpha[\tau]=\boldsymbol\alpha[\tau] \\
& \Rightarrow \\
\dot{a}(\tau) &=\dot{\mathbf{a}}(\tau)=\left(Q_{a a}-\Gamma_{a}\right) \nabla_{a} F(\pi)+\ldots \\
\dot{\mu}(\tau) &=\dot{\boldsymbol{\mu}}(\tau)=\left(Q_{\mu \mu}-\Gamma_{\mu}\right) \nabla_{\mu} F(\pi)+\ldots
\end{aligned}
\end{equation}	

 \begin{figure}[t!]
    \centering
    \includegraphics[width=0.9\textwidth]{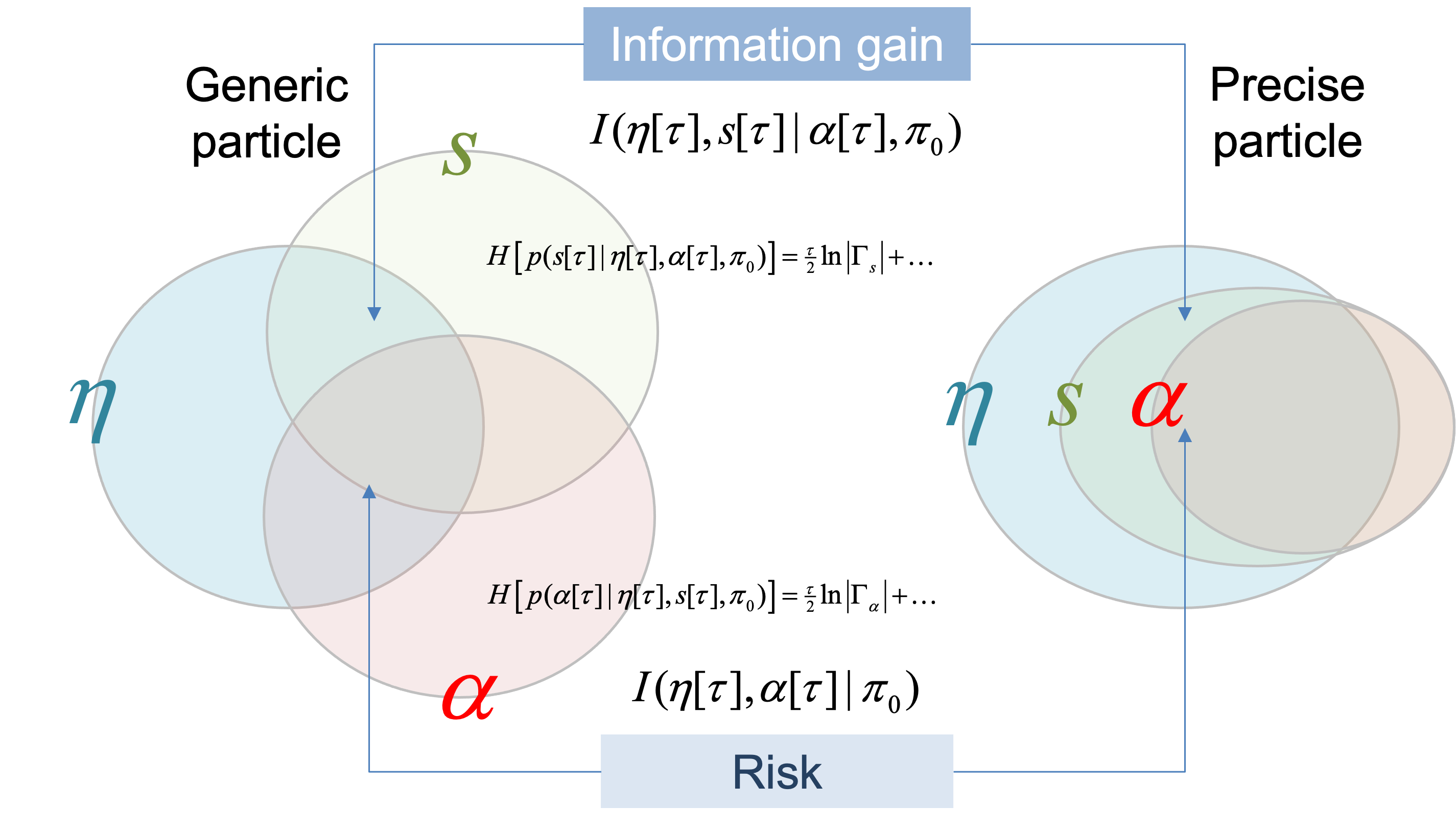}
    \caption{\textbf{Generic and precise particles}. These information diagrams depict the entropy of external, sensory and autonomous paths, where intersections correspond to shared or mutual information. A conditional entropy corresponds to an area that is outside the variable upon which the entropy is conditioned. The diagram on the left shows the generic case, in which uncertainty about paths inherits from random fluctuations that determine the conditional entropies of paths. When the amplitude of random fluctuations on the motion of particular states is very small, we have precise particles in which there is no uncertainty about autonomous paths, given sensory paths (the right information diagram). Similarly, there is no uncertainty about sensory paths given external and autonomous paths. Note that because we are dealing with continuous states, we are implicitly interpreting the entropies as the limiting density of discrete points (LDDP), which have a lower bound of zero \cite{jaynesInformationTheoryStatistical1957}. (LDDP is an adjustment to differential entropy which ensures that entropy is lower bounded by zero. LDDP equals the negative KL-divergence between the density in question and a uniform density). Two relative entropies (information gain and risk) are highlighted as areas of intersection. These will play an important role later, when decomposing the action (i.e., expected free energy) of autonomous paths.}
    \label{fig: 4}
\end{figure}

This suggests that precise particles—such as you and me—will respond to environmental flows and fluctuations in a precise and predictable fashion. Figure \ref{fig: 4} illustrates the difference between generic and precise particles using an information diagram. Note that for precise particles, there is no uncertainty about autonomous states, given sensory states. This follows because the flow of autonomous states depends only on sensory states and themselves. Is the behaviour of precise particles a sufficient description of sentient behaviour? 

\begin{figure}[t!]
    \centering
    \includegraphics[width=\textwidth]{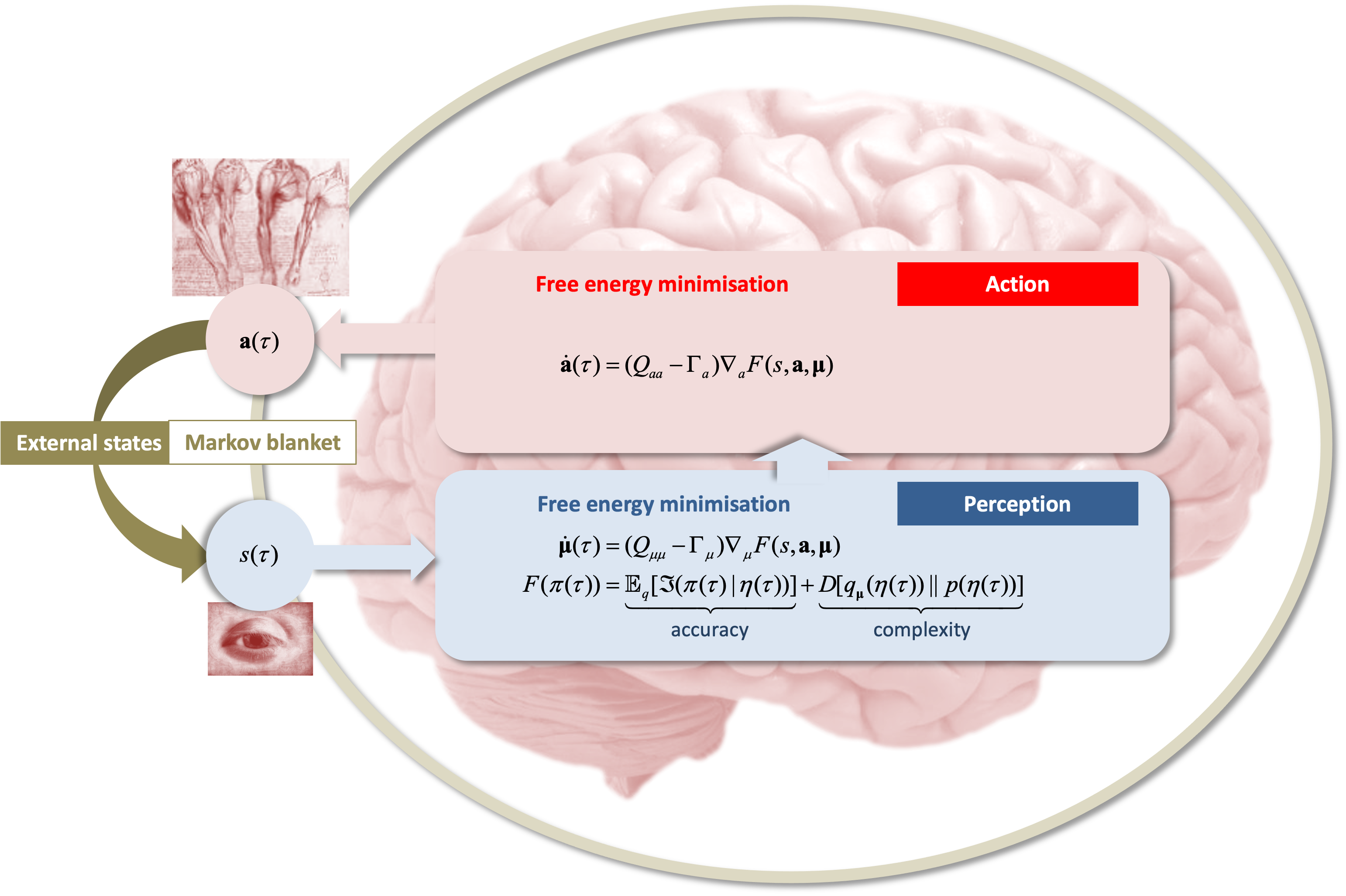}
    \caption{\textbf{Bayesian mechanics and active inference}. This graphic summarises the belief updating implicit in gradient flows on variational free energy. These are the paths taken by a precise particle or the paths of least action of a generic particle. It illustrates a simple form of (active) inference that has been used in a variety of applications and simulations; ranging from handwriting and action observation~\cite{fristonActionUnderstandingActive2011}, through to birdsong and generalised synchrony in communication~\cite{fristonActiveInferenceCommunication2015}. In brief, sensory states furnish free energy gradients (often expressed as prediction errors), under some generative model. Neuronal dynamics are simulated as a flow on the resulting gradients to produce internal states that parameterise posterior beliefs about external states. Similarly, active states are simulated as a flow on free energy gradients that generally play the role of prediction errors. In other words, active states mediate motor or autonomic reflexes~\cite{feldmanNewInsightsActionperception2009,mansellControlPerceptionShould2011}. An example of this kind of active inference is provided in the next figure.}
    \label{fig: 5}
\end{figure}

\begin{figure}
    \centering
    \includegraphics[width=0.8\textwidth]{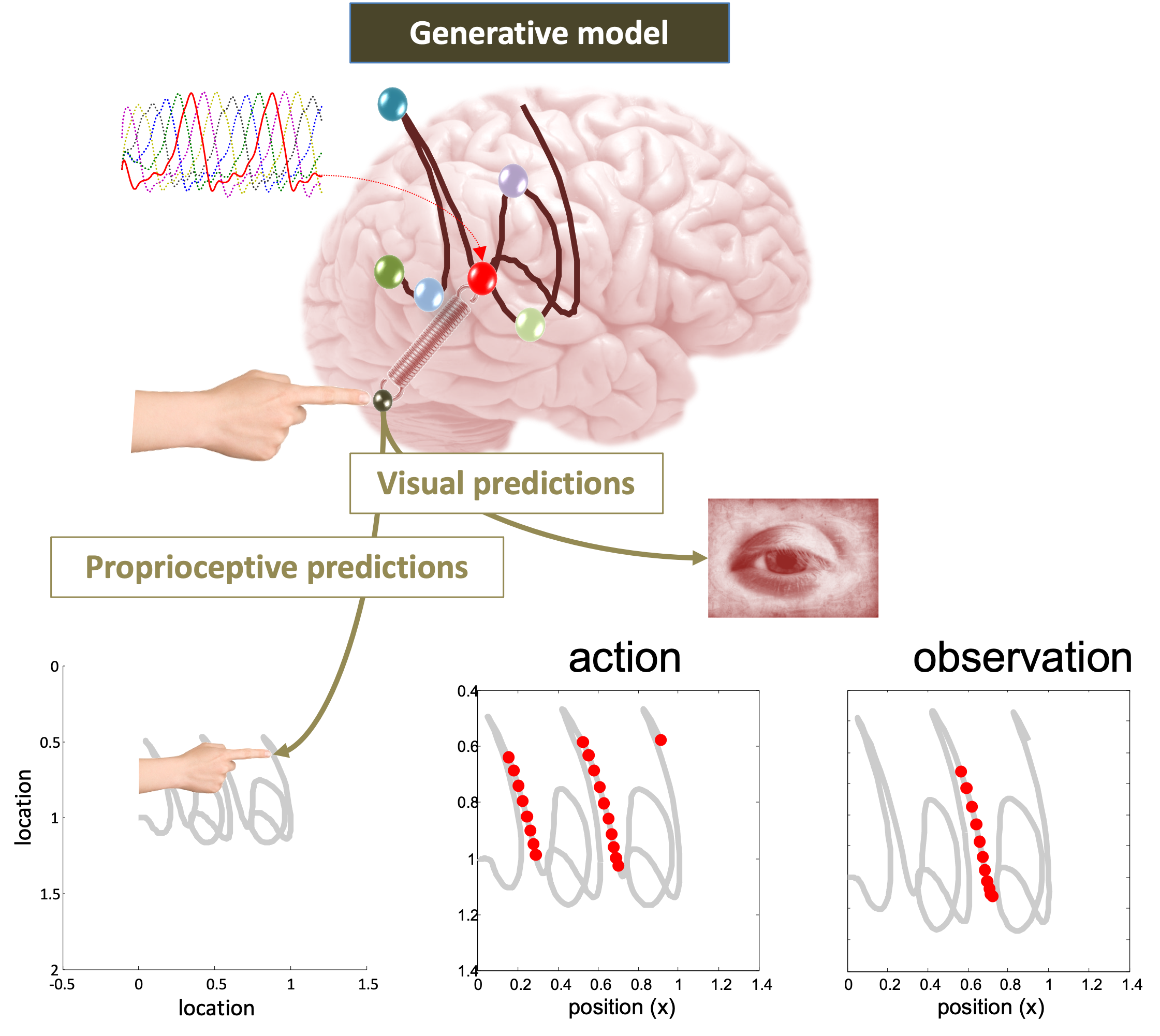}
    \caption{\textbf{Sentient behaviour and action observation}. This figure illustrates a simulation of active inference (here, writing) evinced by a precise particle, in terms of inferences about external states of the world, consequent predictions about sensory input, and ensuing action. The autonomous dynamics that underwrite this behaviour rest upon a generative model of sensory states in the form of Lotka-Volterra dynamics; see sample sensory trajectories as (arbitrarily) coloured lines in the upper left inset. The generative model defines the joint density under which internal trajectories can be seen as parameterising external states. This model is not a description of the true external states (which here are simply the positions of the joints in the simulated arm---with dynamics given by simple Newtonian rules). In this generative model, external trajectories are  assumed to follow predator-prey like dynamics such that a succession of peaks are generated for a subset of external states (or coordinates) in turn. Each coordinate is associated with a location in Euclidean space that attracts the agent’s finger (the active states); i.e., with a trajectory towards that attracting point. The resulting attracting point is thus a weighted sum of each possible attracting point weighted by the coordinates following the Lotka-Volterra trajectory. In turn, the internal states supply predictions of what sensory states should register if the agent’s beliefs were true. Active states (i.e., the forces driving changes in the angular velocities of the limb joints) try to suppress the ensuing prediction error by adjusting expected changes in sensed angular velocity, through exerting forces on the agent’s joints (not shown). The subsequent movement of the arm is traced out in the lower-left panel. This trajectory has been plotted in a moving frame of reference so that it looks like handwriting (e.g., a succession of ‘j’ and ‘a’ letters). The lower right panels show the activity of one internal state during distinct phases of ‘action’, and ‘action-observation’. During the action phase, sensory states register the visual and proprioceptive consequences of movement, while under action observation, only visual sensations are available—as if the agent was watching another agent. The red dots correspond to the times during which this internal state exceeded an arbitrary threshold. The key thing to note here is that this internal state responds preferentially when, and only when, the motor trajectory produces a down-stroke, but not an up-stroke—evincing a cardinal feature of neuronal responses, namely, their functional selectivity. Furthermore, with a slight delay, this internal state responds during action and action observation. From a biological perspective, this is interesting because it speaks to an empirical phenomenon known as mirror neuron activity~\cite{galleseMirrorNeuronsSimulation1998,rizzolattiMirrorneuronSystem2004,kilnerPredictiveCodingAccount2007}. Please see~\cite{fristonActionUnderstandingActive2011} for further details.}
    \label{fig: 6}
\end{figure}

On one reading, perhaps: one can reproduce biological behaviour by numerically integrating \eqref{eq: 21} or \eqref{eq: 27} under a suitable generative (state-space) model specifying the motion of external and sensory states\footnote{A generative model can be specified through the flow of external or sensory states, and the random fluctuations of their motion; that is, the first two lines of \eqref{eq: 27}. Observing that the free energy \eqref{eq: 26} is only a function of these flows and the covariance of fluctuations, it is sufficient to specify those covariances, rather than the whole structure of the fluctuations.}. Figure \ref{fig: 5} illustrates the implicit computational architecture used to simulate sentient behaviour by integrating \eqref{eq: 21}. This scheme allows one to simulate the internal and active states through sensory states caused by external dynamics. Figure \ref{fig: 6} showcases an example from the active inference literature, that integrates \eqref{eq: 27} under a suitably specified generative model, to simulate sentient behaviour that looks like handwriting. The details of the simulation and the details of the generative model are not relevant here but are summarised in the figure legend; what is important is to get a sense of the kind of behaviour that can be reproduced by integrating \eqref{eq: 27}.

The example in Figure \ref{fig: 6} illustrates an application of the free energy principle. Here, instead of describing a system by deriving its NESS density, we have specified some equations of motion (and covariance of random fluctuations) to realise particular dynamics using \eqref{eq: 27} and \eqref{eq: free energy generalised}. In effect, we have simulated self-evidencing, starting from a definition (i.e., state-space generative model) of paths that characterise this kind of particle\footnote{The example in Figure \ref{eq: 6} used \eqref{eq: 27} with generalised coordinates of motion up to fourth order. 
Numerical analyses suggest that simulating generalised motion up to order six (i.e., ignoring all subsequent orders of motions) is sufficient in most circumstances \cite{fristonVariationalTreatmentDynamic2008}.}.

These simulations speak to a key aspect in the applications of the FEP. Hitherto, we have simply defined the variational density as the conditional density over external states given a sensory state. However, when simulating precise particles through a gradient flow on variational free energy, as in \eqref{eq: 21} or \eqref{eq: 27}, the requisite gradients have to be evaluated. In turn, this requires the functional form of the variational density or posterior distribution, which may be difficult to compute exactly\footnote{In Bayesian inference, it is well-known that computing the posterior distribution given data and a generative model $p(\eta \mid \pi) = p(\eta, \pi)/p(\pi)$ is computationally costly as it involves computing a (typically) high-dimensional integral $p(\pi)=\int p(\eta, \pi) d \eta$ (i.e., a partition function).}. In this case, we take a variational density that approximates the true posterior, whence the variational free energy becomes an upper bound on surprisal: see \eqref{eq: free energy upper bound}. From the perspective of Bayesian inference, this takes us from (computationally costly) \textit{exact} Bayesian inference to (computationally cheap) \textit{approximate} Bayesian inference~\cite{bealVariationalAlgorithmsApproximate2003,winnVariationalMessagePassing2005,dauwelsVariationalMessagePassing2007}. On one reading of its inception, this is why variational free energy was introduced \cite{feynmanStatisticalMechanicsSet1998}; namely, to convert a computationally expensive marginalisation problem into a computationally manageable optimisation problem. Note that when using generalised coordinates to realise active inference; i.e., \eqref{eq: 27}, we are generally employing approximate Bayesian inference: the functional form of the variational density inherits directly from Gaussian assumptions about random fluctuations, however the expansion in generalised coordinates on which it is based upon \eqref{eq: 24} is generally an approximation to the underlying dynamic (cf. \ref{footnote expansion in gen coords}).

\subsection{Summary}
Precise particles, immersed in an imprecise world, respond (almost) deterministically to external fluctuations\footnote{\textbf{Question}: does the absence of random fluctuations preclude dissipative gradient flows? No, because the gradients can increase with the precision of random fluctuations. In the limit of no random fluctuations, the steady-state density tends towards a delta function (i.e., a fixed-point attractor) and the dissipative gradients tend towards infinity.}. This means, given a generative model (i.e., NESS density), one can solve the equations of motion in \eqref{eq: 27} to predict how autonomous states evolve as they pursue their path of least action. So, why might this limiting behaviour be characteristically biological?

Precise particles may be the kind of particles that show lifelike or biotic behaviour, in the sense they respond predictably, given their initial states and the history of external influences. The distinction between imprecise (e.g., statistical) and precise (e.g., classical) particles rests on the relative contribution of dissipative and conservative flow to their path through state-space, where solenoidal flow predominates in the precise setting. This means precise particles exhibit solenoidal behaviour such as oscillatory and (quasi) periodic orbits—and an accompanying loss of detailed balance, i.e., turbulent and time-irreversible dynamics~\cite{andresMotionSpikesTurbulentLike2018,decoTurbulentlikeDynamicsHuman2020,dacostaEntropyProductionStationary2022}. On this view, one might associate precise particles with living systems with characteristic biorhythms~\cite{lopesdasilvaNeuralMechanismsUnderlying1991,kopellNeuronalAssemblyDynamics2011,arnalCorticalOscillationsSensory2012,buzsakiScalingBrainSize2013}; ranging from gamma oscillations in neuronal populations, through slower respiratory and diurnal cycles to, perhaps, lifecycles per se. Turning this on its head, one can argue that living systems are a certain kind of particle that, in virtue of being precise, evince conservative dynamics, biorhythms and time irreversibility.

One might ask if solenoidal flow confounds the gradient flows that underwrite self-evidencing. In fact, solenoidal flow generally augments gradient flows—or at least this is what it looks like. In brief, the mixing afforded by solenoidal flow can render gradient descent more efficient~\cite{ottControllingChaos1990,hwangAcceleratingDiffusions2005,hwangAcceleratingGaussianDiffusions1993,lelievreOptimalNonreversibleLinear2013,aslimaniNewHybridAlgorithm2018}. An intuitive example is stirring sugar into coffee. The mixing afforded by the solenoidal stirring facilitates the dispersion of the sugar molecules down their concentration gradients. On this view, the solenoidal flow can be regarded as circumnavigating the contours of the steady-state density to find a path of steepest descent.

The emerging picture here is that biotic systems feature solenoidal flow, in virtue of being sufficiently large to average away random fluctuations, when coarse-graining their dynamics \cite{fristonFreeEnergyPrinciple2019a}. From the perspective of the information geometry induced by the FEP, this means biological behaviour may be characterised by internal solenoidal flows that do not change variational free energy—or surprisal—and yet move on the internal (statistical) manifold to continually update Bayesian beliefs about external states. Biologically, this may be a description of central pattern generators \cite{arnalCorticalOscillationsSensory2012,grossSpeechRhythmsMultiplexed2013} that underwrite rhythmical activity (e.g., walking and talking) that is characteristic of biological systems \cite{buzsakiNeuronalOscillationsCortical2004}. The example in Figure \ref{fig: 6} was chosen to showcase the role of solenoidal flows in Bayesian mechanics that—in this example—arise from the use of Lotka-Volterra dynamics in the generative model. In psychology, this kind of conservative active inference may be the homologue of being in a ‘flow state’~\cite{csikszentmihalyiFlowPsychologyOptimal2008}.

In short, precise particles may be the kind of particles we associate with living systems. And precise particles have low entropy paths. If so, the question now becomes: what long-term behaviour does this class of particle show? In other words, instead of asking which behaviours \textit{lead} to low entropy dynamics, we can now ask which behaviours \textit{follow from} low entropy dynamics? We will see next that precise particles appear to plan their actions and, perhaps more interestingly, show information and goal-seeking behaviour.

\section{Path integrals, planning and curious particles}
\label{sec: curious}

While the handwriting example in Figure \ref{fig: 6} offers a compelling simulation of self-evidencing—in the sense of an artefact creating its own sensorium—there is something missing as a complete account of sentient behaviour. This is because we have only considered the response of autonomous states to sensory states over limited periods of time. To disclose a deeper Bayesian mechanics, we need to consider the paths of autonomous states over extended periods. This takes us to the final step and back to the path-integral formulation.

In the previous section, we focused on linking dynamics to densities over (generalised) states. In brief, we saw that internal states can be construed as parameterising (Bayesian) beliefs about external states at any point in time. In what follows, we move from densities over \textit{states} to densities over \textit{paths}—to characterise the behaviour of particles in terms of their trajectories.

In what follows, we will be dealing with predictive posterior densities over external and particular paths, given (initial) particular states, which can be expressed in terms of the variational density parameterised by the current (initial) internal state:\footnote{\textbf{Question:} Why is the variational density parameterised by the initial internal state rather than the initial internal mode? The answer is that in precise particles, the absence of fluctuations on particular dynamics means that the internal states always coincides with the internal mode.}
 \begin{equation}
 \label{eq: 29}
\begin{aligned}
q\left(\eta[\tau], \pi[\tau] \mid \pi_{0}\right) & \triangleq \mathbb{E}_{q_{\mu}}\left[p\left(\eta[\tau], \pi[\tau] \mid \eta_{0}, \pi_{0}\right)\right]=p\left(\eta[\tau], \pi[\tau] \mid \pi_{0}\right) \\
q_{\mu}\left(\eta_{0}\right) &=p\left(\eta_{0} \mid \pi_{0}\right).
\end{aligned}
\end{equation}	

All this equation says is that, given the initial particular states, we can evaluate the joint density over external and particular paths, because we know the density over the initial external states, which is parameterised by the initial internal state.

We are interested in characterising autonomous responses to initial particular states. This is given by the action of autonomous paths as a function of particular states. In other words, we seek an expression for the probability of an autonomous path that \textit{(i)} furnishes a teleological description of self-organisation and \textit{(ii)} allows us to simulate the sentient trajectories of particles, given their sensory streams. Getting from the action of particular paths to the action of autonomous paths requires a marginalisation over sensory paths. This is where the precise particle assumption comes in: it allows us to eschew this (computationally costly) marginalisation by expressing the action of particular paths as an \textit{expected free energy}.

Recall that when random fluctuations on the motion of particular states vanish, there is no uncertainty about autonomous paths, given external and sensory paths. And there is no uncertainty about sensory paths given external and autonomous paths. If we interpret entropies as the limiting density of discrete points (see Figure \ref{fig: 4}), then the uncertainty about particular, autonomous and sensory paths, given external paths, become interchangeable:
 \begin{equation}
 \label{eq: 30}
\begin{aligned}
\lim _{\Gamma_{\pi} \rightarrow 0} \H\left[p\left(\pi[\tau] \mid \eta[\tau], \pi_{0}\right)\right] &=\underbrace{\H\left[p\left(\alpha[\tau] \mid \eta[\tau], s[\tau], \pi_{0}\right)\right]}_{=0}+\H\left[p\left(s[\tau] \mid \eta[\tau], \pi_{0}\right)\right] \\
&=\underbrace{\H\left[p\left(s[\tau] \mid \eta[\tau], \alpha[\tau], \pi_{0}\right)\right]}_{=0}+\H\left[p\left(\alpha[\tau] \mid \eta[\tau], \pi_{0}\right)\right] \\
& \Rightarrow \\
\mathbb{E}_{q}\left[\ln p\left(\pi[\tau] \mid \eta[\tau], \pi_{0}\right)\right] &=\mathbb{E}_{q}\left[\ln p\left(s[\tau] \mid \eta[\tau], \pi_{0}\right)\right]=\mathbb{E}_{q}\left[\ln p\left(\alpha[\tau] \mid \eta[\tau], \pi_{0}\right)\right]
\end{aligned}
\end{equation}
We can leverage this exchangeability to express the action of autonomous paths in terms of an expected free energy. From \eqref{eq: 29} and \eqref{eq: 30}, we have (dropping the conditioning on initial states for clarity):
 \begin{equation}
 \label{eq: 31}
\begin{aligned}
&0=\mathbb{E}_{q}\left[\ln \frac{p(\eta[\tau], \alpha[\tau])}{q(\eta[\tau], \alpha[\tau])}\right]=\mathbb{E}_{q}\left[\ln \frac{p(\alpha[\tau] \mid \eta[\tau]) p(\eta[\tau])}{q(\eta[\tau] \mid \alpha[\tau]) q(\alpha[\tau])}\right]\\
&=\mathbb{E}_{q}\left[\ln \frac{p(s[\tau] \mid \eta[\tau]) p(\eta[\tau])}{q(\eta[\tau] \mid \alpha[\tau])}-\ln q(\alpha[\tau])\right]=\mathbb{E}_{q(\alpha[\tau])}[\mathcal{A}(\alpha[\tau])-G(\alpha[\tau])]\\
&=D\left[q(\alpha[\tau]) \| \mathrm{e}^{-G}\right] \Rightarrow G(\alpha[\tau])=\mathcal{A}(\alpha[\tau])\\
&G(\alpha[\tau])=\mathbb{E}_{q(\eta[\tau], s[\tau] \alpha[\tau])}[\overbrace{\underbrace{\ln q(\eta[\tau] \mid \alpha[\tau])-\ln p(\eta[\tau])}_{\text {Expected complexity }}}^{\text {Risk }} \overbrace{-\underbrace{\ln p(s[\tau] \mid \eta[\tau])}_{\text {Expected accuracy }}}^{\text {Ambiguity }}]
\end{aligned}
\end{equation}	
All we have done here is to exchange the density over autonomous paths, conditioned on external paths, with the corresponding density over sensory paths (in the second line) thanks to the precise particle assumption. By gathering terms into a functional of autonomous paths, we recover autonomous action as an expected free energy.

 \begin{figure}
    \centering
    \includegraphics[width=\textwidth]{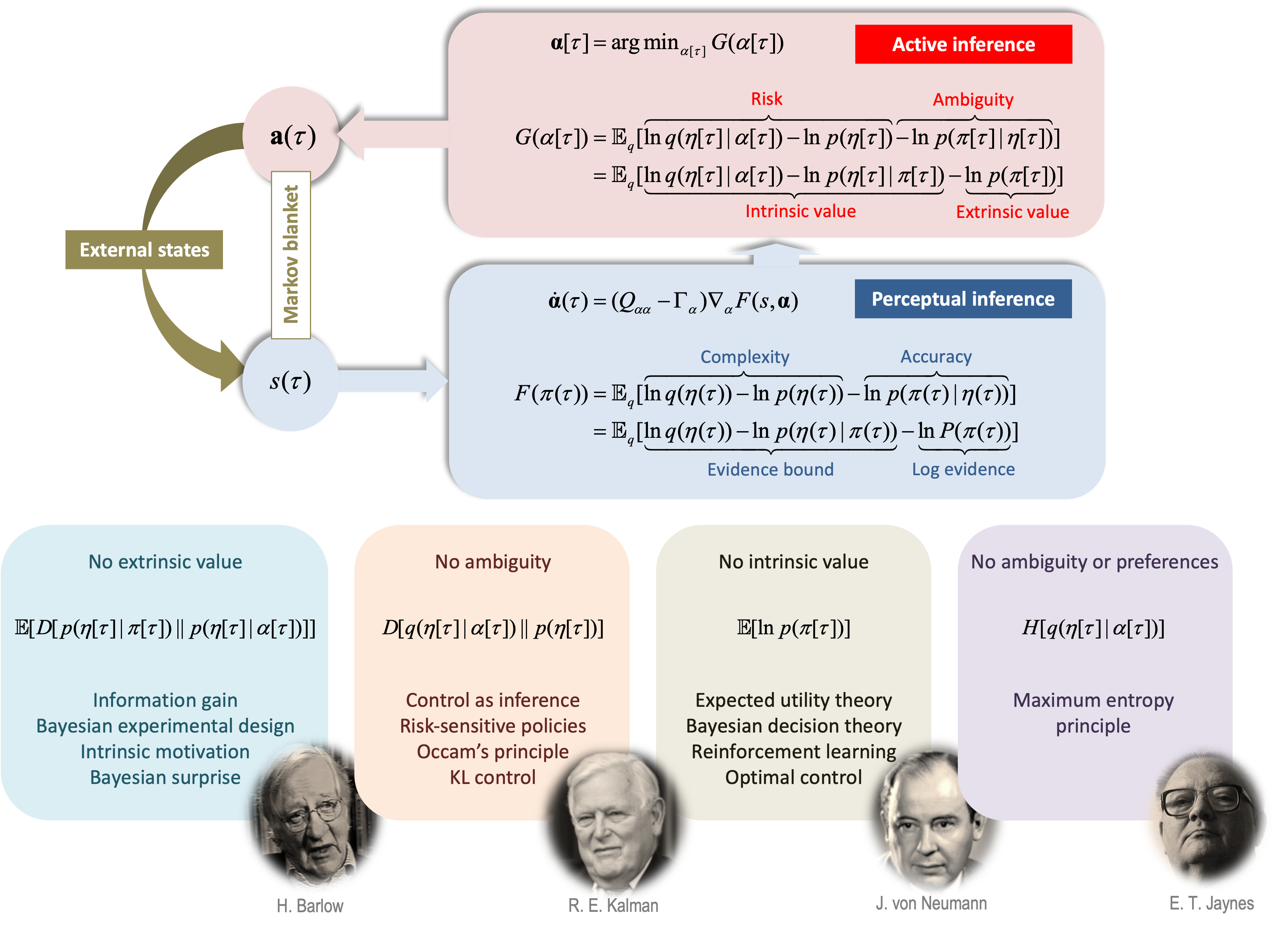}
    \caption{\textbf{Expected free energy and active inference}. This figure illustrates active inference, and highlights various points of contact with other accounts of sentient, purposeful or intelligent behaviour. The upper panel casts action and perception as the minimisation of expected and variational free energy, respectively. Crucially, the path integral formulation of active inference introduces posterior beliefs over autonomous paths (i.e., policies) that entail a description of planning as inference~\cite{attiasPlanningProbabilisticInference2003,botvinickPlanningInference2012,kaplanPlanningNavigationActive2018}. When simulating active inference, posterior beliefs about external paths, under plausible policies, are optimised by a gradient flow on the variational (free energy) bound on log evidence—as in Figure \ref{fig: 3}. These beliefs are then used to evaluate the expected free energy of allowable policies, from which actions can be selected~\cite{fristonActiveInferenceProcess2017,dacostaActiveInferenceDiscrete2020,barpGeometricMethodsSampling2022a}. Crucially, expected free energy contains terms that arise in various formulations of optimal behaviour that predominate in cognitive science, engineering and economics. These terms are disclosed when one removes certain sources of uncertainty.
    For example, if we remove ambiguity, decision-making minimises risk, which corresponds to aligning predictions with preferences about the external course of events. This underwrites prospect theory of human choice behaviour in economics~\cite{kahnemanProspectTheoryAnalysis1979} and modern approaches to control as inference~\cite{levineReinforcementLearningControl2018,rawlikStochasticOptimalControl2013,toussaintRobotTrajectoryOptimization2009}, variously known as Kalman duality~\cite{kalmanNewApproachLinear1960,todorovGeneralDualityOptimal2008a}, KL control~\cite{kappenOptimalControlGraphical2012} and maximum entropy reinforcement learning~\cite{ziebartModelingPurposefulAdaptive2010}.
    If we further remove preferences, decision-making maximises the entropy of external trajectories. This maximum entropy principle~\cite{jaynesInformationTheoryStatistical1957,lasotaChaosFractalsNoise1994} can be interpreted as least committing to a presupposed external trajectory and therefore keeping options open \cite{klyubinKeepYourOptions2008}.
    If we reintroduce ambiguity, but ignore preferences, decision-making maximises intrinsic value or expected information gain~\cite{mackayInformationTheoryInference2003}. This underwrites Bayesian experimental design~\cite{lindleyMeasureInformationProvided1956} and active learning in statistics~\cite{mackayInformationBasedObjectiveFunctions1992}, intrinsic motivation and artificial curiosity in machine learning and robotics~\cite{oudeyerWhatIntrinsicMotivation2007,schmidhuberFormalTheoryCreativity2010,bartoNoveltySurprise2013,sunPlanningBeSurprised2011,deciIntrinsicMotivationSelfDetermination1985}. This is mathematically equivalent to optimising expected Bayesian surprise and mutual information, which underwrites visual search~\cite{ittiBayesianSurpriseAttracts2009,parrGenerativeModelsActive2021} and the organisation of our visual apparatus~\cite{barlowPossiblePrinciplesUnderlying1961,linskerPerceptualNeuralOrganization1990,opticanTemporalEncodingTwodimensional1987a}.
    Lastly, if we remove intrinsic value, we are left with maximising extrinsic value or expected utility, which underwrites expected utility theory~\cite{vonneumannTheoryGamesEconomic1944}, game theory, optimal control~\cite{bellmanDynamicProgramming1957, astromOptimalControlMarkov1965a} and reinforcement learning~\cite{bartoReinforcementLearningIntroduction1992}. Bayesian formulations of maximising expected utility under uncertainty are also known as Bayesian decision theory~\cite{bergerStatisticalDecisionTheory1985}.
    The expressions for variational and expected free energy are arranged to illustrate the relationship between \textit{complexity} and \textit{accuracy}, which become \textit{risk} and \textit{ambiguity} in the path integral formulation. This suggests that risk-averse policies minimise expected complexity or computational cost~\cite{schmidhuberFormalTheoryCreativity2010}.}
    \label{fig: 7}
\end{figure}

By analogy with the expression for variational free energy \eqref{eq: 18}, the expressions for the expected free energy in \eqref{eq: 31} suggest that \textit{accuracy} becomes \textit{ambiguity}, while \textit{complexity} becomes \textit{risk}. So why have we called these terms ambiguity and risk? Ambiguity is just the expected precision or conditional uncertainty about sensory states given external states. A heuristic example of an imprecise likelihood mapping—between external and sensory paths—would be a dark room, where there is no precise information at hand. Indeed, according to \eqref{eq: 31}, sensory paths into dark rooms should be highly unlikely. However, this is not the complete story, in the sense that the risk puts certain constraints on any manifest tendency to minimise ambiguity. 

Here, risk is simply the divergence between external paths given an autonomous path (i.e., policy or plan), relative to external states of affairs. The marginal density over external paths is often referred to in terms of \textit{prior preferences}, because they constitute the priors of the generative model characterising the particle's behaviour \cite{parrGeneralisedFreeEnergy2019}. In short, the expression for expected free energy, suggests that particles will look as if they are (i) minimising the risk of incurring external trajectories that diverge from prior preferences, while (ii) resolving ambiguity in response to external events. In this formulation, autonomous paths play the dual role of registering the influences of external events (via ambiguity), while also causing those events (via risk). 

The autonomous path with the least expected free energy is the most likely path taken by the autonomous states. 
 \begin{equation}
 \label{eq: 32}
\begin{aligned}
G(\alpha[\tau]) &=\mathcal{A}(\alpha[\tau]) \\
& \Rightarrow \boldsymbol \alpha[\tau]=\arg \min _{\alpha[\tau]} G(\alpha[\tau]) \\
& \Rightarrow \delta_{\alpha} G(\boldsymbol \alpha[\tau])=0 \\
\mathbb{E}[G(\alpha[\tau])] &=\mathbb{E}[\mathcal{A}(\alpha[\tau])]=\H[p(\alpha[\tau])]
\end{aligned}
\end{equation}	

In short, expected free energy scores the autonomous action of particles that do not admit noisy dynamics. Expected free energy has a specific form that inherits from the assumption that the amplitude of particular fluctuations is small, which is the case for precise articles by definition. Although variational and expected free energy are formally similar, they are fundamentally different kinds of functionals: variational free energy is a functional of a density over states, while expected free energy is a functional of a density over paths. Variational free energy can also be read as a function of particular states, while expected free energy is a function of an autonomous path. Finally, variational free energy is a bound on surprisal, while expected free energy is not a bound—it is the action of autonomous trajectories.

Expected free energy plays a definitive role in active inference, where it can be regarded as a fairly universal objective function for selecting autonomous paths of least action. Figure \ref{fig: 7} shows that the expected free energy contains terms that arise in various formulations of optimal behaviour; ranging from optimal Bayesian design~\cite{lindleyMeasureInformationProvided1956} through to control as inference~\cite{ziebartModelingPurposefulAdaptive2010,levineReinforcementLearningControl2018}. We refer the reader to ~\cite{dacostaRewardMaximizationDiscrete2023,fristonSophisticatedInference2021,sajidActiveInferenceBayesian2022,hafnerActionPerceptionDivergence2020,millidgeRelationshipActiveInference2020a} for formal investigations of the relationship between these formulations.

\begin{figure}[t!]
    \centering
    \includegraphics[width=\textwidth]{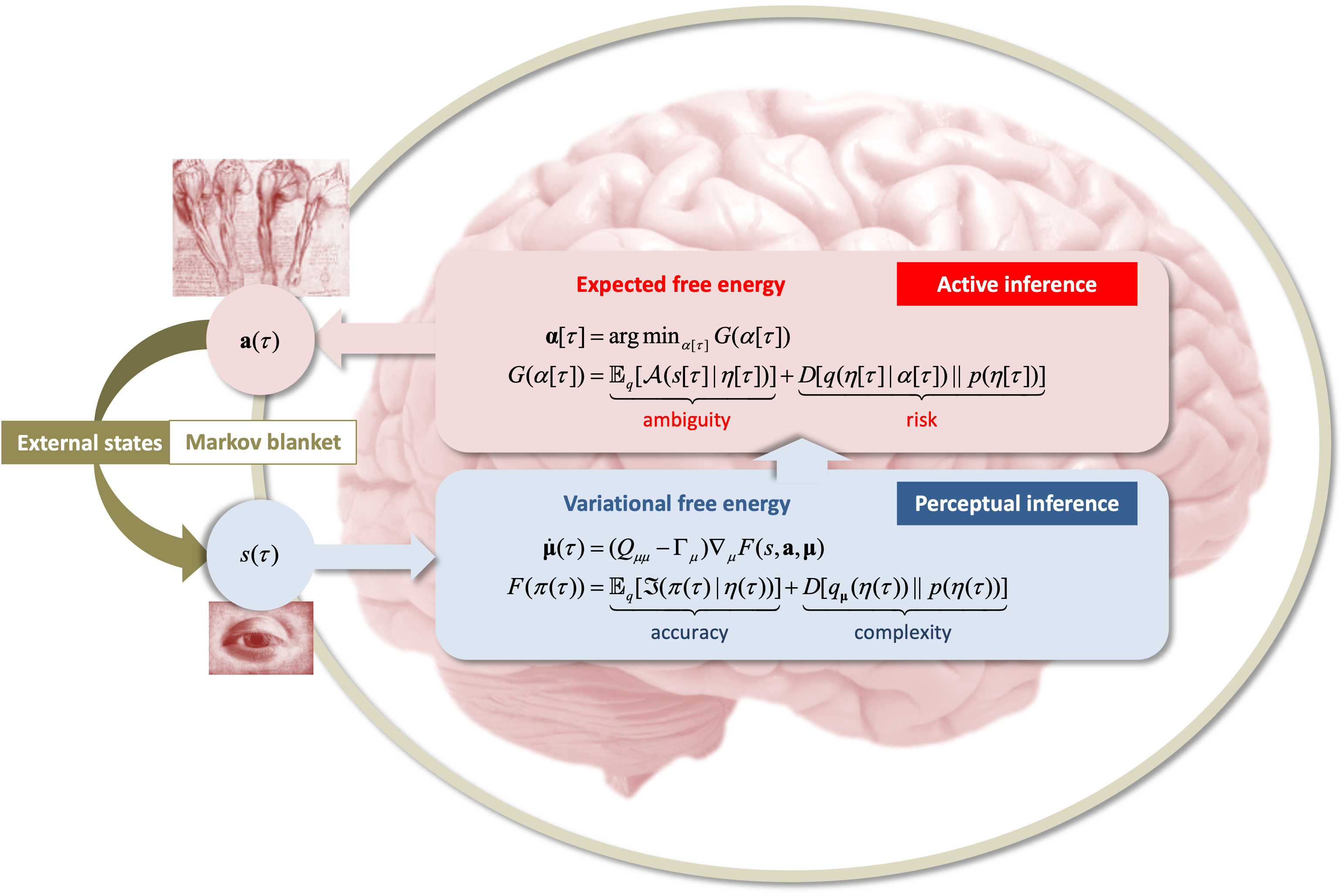}
    \caption{\textbf{Bayesian mechanics and active inference}. This graphic summarises the belief updating implicit in the minimisation of variational and expected free energy. It describes active inference based upon autonomous paths or policies and has been used in a variety of applications and simulations; ranging from games in behavioural economics~\cite{fitzgeraldActiveInferenceEvidence2015} and reinforcement learning~\cite{schwartenbeckEvidenceSurpriseMinimization2015,sajidActiveInferenceDemystified2021} through to language understanding~\cite{fristonDeepTemporalModels2018} and scene construction~\cite{mirzaSceneConstructionVisual2016}. In this setup, actions solicit a sensory outcome that informs approximate posterior beliefs about hidden or external states of the world—via minimisation of variational free energy under a set of plausible policies (i.e., perceptual inference). The approximate posterior beliefs are then used to evaluate expected free energy and subsequent action (i.e., active inference). A key insight from simulations is that the form of the generative model can be quite different from the process by which external states generate sensory states. In effect, this enables agents (i.e., particles) to author their own sensorium in a fashion that has close connections with econiche construction~\cite{bruinebergSelforganizationFreeEnergy2014}. Please see~\cite{fristonGraphicalBrainBelief2017,dacostaActiveInferenceDiscrete2020} for technical details and for a heuristic discussion of how the belief updating could be implemented in the brain.}
    \label{fig: 8}
\end{figure}

\begin{figure}
    \centering
    \includegraphics[width=0.85\textwidth]{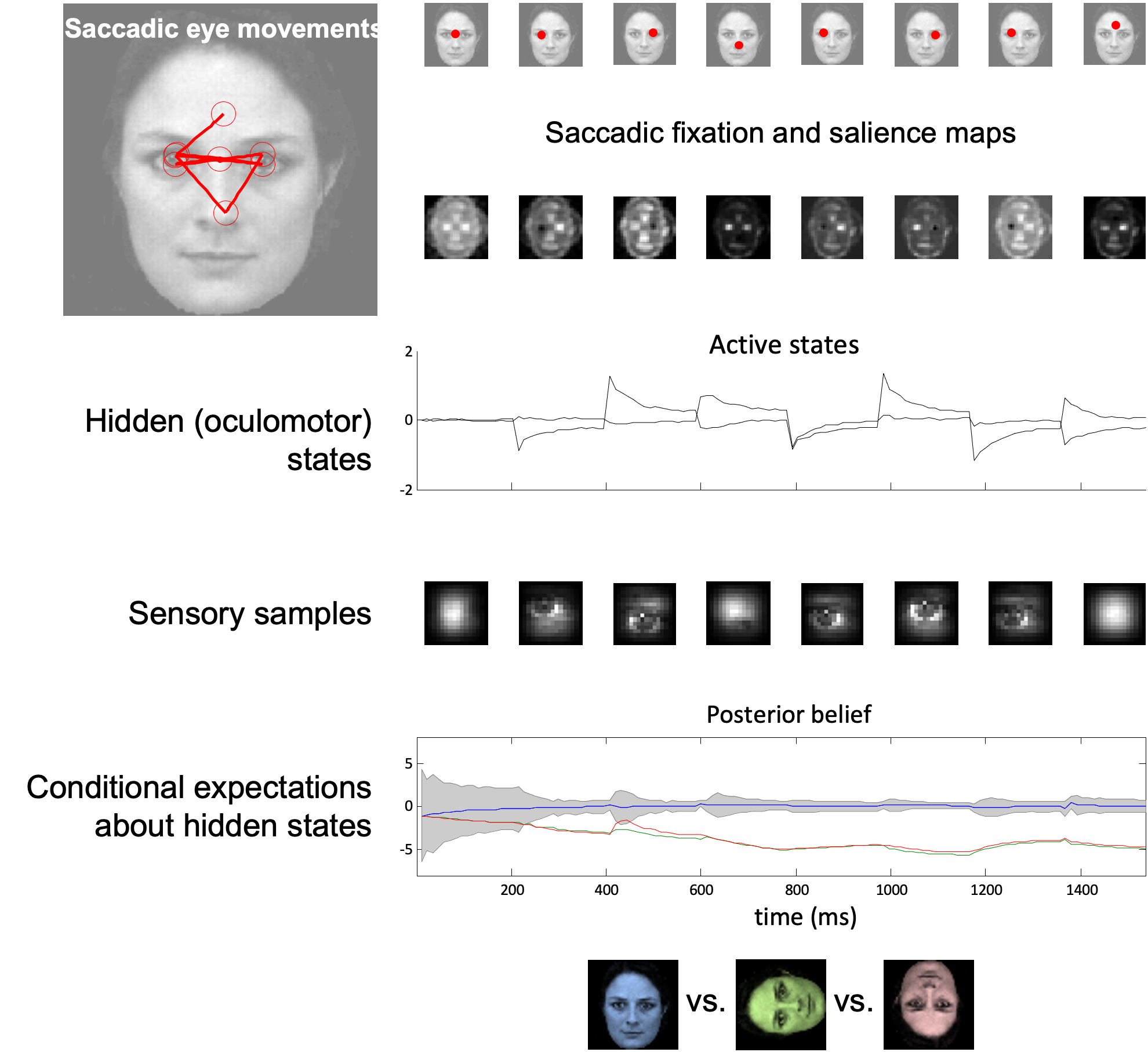}
    \caption{\textbf{Epistemic foraging}. This figure shows the results of a numerical simulation where a face was presented to an agent, whose responses were obtained by selecting active states that minimised expected free energy following an eye movement. The agent had three internal images or hypotheses (i.e., internal states) about the external state she might sample (an upright face (blue), an inverted face (magenta) and a rotated face (green)---shown at the bottom). The agent was presented with sensory samples of an upright face and her variational posterior over the external state was obtained by descending variational free energy over a 12ms time bin until the next saccade (i.e., action) was emitted. This perception-action cycle was repeated eight times. The agent's eye movements are shown as red dots at the end of each saccade in the upper row. The corresponding sequence of eye movements is shown in the upper-left inset, where the red circles correspond roughly to the proportion of the visual image sampled. These saccades are driven by the salience maps in the second row, which correspond to the expected free energy as a function of the policies; namely, the next saccade or where to look next. As expected free energies are defined in terms of trajectories, it is best to see the locations of on these salience maps as expressing the expected free energy of a trajectory that ends in that location. Note that these maps change with successive saccades as variational posterior beliefs become progressively more confident about the external state. Note also that salience is depleted in locations that were foveated in the previous saccade because these locations no longer have epistemic affordance or expected information gain (i.e., the ability to reduce uncertainty in the expected free energy). In neuroscience, this empirical phenomenon is known as inhibition of return. Oculomotor responses are shown in the third row in terms of the two oculomotor states corresponding to vertical and horizontal eye movements. The associated portions of the image sampled (at the end of each saccade) are shown in the fourth row. The fifth row shows the evolution of variational posterior beliefs about external (a.k.a. hidden) in terms of the log probability they assign to each possible external state (colour coded) and 90\% confidence intervals. The key thing to note is that the credence about the true external state supervenes over alternative expectations and, as a result, confidence about the category increases (and confidence intervals shrink to the mode). This illustrates the nature of evidence accumulation when selecting a hypothesis or percept that best explains sensory states. Please see~\cite{fristonPerceptionsHypothesesSaccades2012} for further details.}
    \label{fig: 9}
\end{figure}

Equipped with a specification of the most likely autonomous path—in terms of expected free energy—we can simulate fairly lifelike behaviour, given a suitable generative model. An example is provided in Figure \ref{fig: 9}---relying upon the computational architecture in Figure \ref{fig: 8}---which illustrates the ambiguity resolving part of the expected free energy in a simulation of visual epistemic foraging.

This epistemic aspect of expected free energy can be seen more clearly if we replace the conditional uncertainty about sensory paths with conditional uncertainty about particular paths, noting that they are the same by \eqref{eq: 30}. After rearrangement, we can express expected free energy in terms of \textit{expected value} and \textit{expected information gain}
~\cite{sajidActiveInferenceBayesian2022,barpGeometricMethodsSampling2022a}:
 \begin{equation}
 \label{eq: 33}
\begin{aligned}
&G(\alpha[\tau])=\mathbb{E}_{q(\eta[\tau], s[\tau] \mid  \alpha[\tau])}[\ln q(\eta[\tau] \mid \alpha[\tau])-\ln p(\eta[\tau])-\ln p(\pi[\tau] \mid \eta[\tau])]\\
&=\underbrace{\overbrace{\mathbb{E}_{q(s[\tau] \mid \alpha[\tau]}[\mathcal{A}(\pi[\tau])]}^{\text {Expected value }}}_{\text {Bayes optimal decisions }}-\underbrace{\overbrace{\mathbb{E}_{q(s[\tau] \mid \alpha[\tau])}\left[D[p(\eta[\tau] \mid s[\tau], \alpha[\tau]) \| p(\eta[\tau] \mid \alpha[\tau])]\right.}^{\text {Expected information gain }}}_{\text {Bayes optimal design}}
\end{aligned}
\end{equation}

This provides a complementary interpretation of expected free energy. The first term can be construed as expected cost in the sense it is the expected action of particular paths. This marginal likelihood scores the plausibility of a particle pursuing this kind of path and is usually interpreted in terms of expected loss (i.e., negative expected reward or utility)~\cite{vonneumannTheoryGamesEconomic1944,bartoReinforcementLearningIntroduction1992}, and pragmatic affordance~\cite{schwartenbeckEvidenceSurpriseMinimization2015,fristonActiveInferenceProcess2017}. The second term corresponds to the expected divergence between posterior beliefs about external paths, given autonomous paths, with and without sensory paths. In other words, it scores the resolution of uncertainty or expected information gain afforded by sensory trajectories arising from a commitment to an autonomous path. In this sense, it is sometimes referred to as epistemic affordance~\cite{fristonGraphicalBrainBelief2017}.

When simulating the kind of planning and active inference afforded by the path integral formulation, one usually works with discrete state-spaces and belief updating over discrete epochs of time~\cite{dacostaActiveInferenceDiscrete2020,fristonActiveInferenceProcess2017}. One can see this as a coarse-graining of continuous space-time into discrete space and time bins, where trajectories of continuous states become sequences of discrete states $x[\tau]=(x_1, \ldots, x_\tau)$. In discrete state-spaces, the generative model is usually formulated as a partially observed Markov decision process \cite{dacostaRewardMaximizationDiscrete2023,dacostaActiveInferenceDiscrete2020,parrActiveInferenceFree2022,smithStepbystepTutorialActive2022}, in which the paths of autonomous states constitute policies, which determine transitions among external states. Plausible policies can then be scored with their expected free energy and the next action is selected from the most likely policy $\alpha=(\alpha_{0}, \ldots, \alpha_{\tau})$\footnote{See \cite{dacostaActiveInferenceDiscrete2020,parrActiveInferenceFree2022} for a derivation of these functional forms in partially observable Markov decision processes.}
 \begin{equation}
 \label{eq: 34}
\begin{aligned}
\mathbf{a} &=\arg \min _{a} G(a, \boldsymbol{\mu}) \\
G &=\mathbb{E}_{Q}\left[\ln Q_{\mu}\left(\eta_{1}, \ldots, \eta_{\tau} \mid \eta_{0}, a\right)-\ln P\left(\eta_{1}, \ldots, \eta_{\tau} \mid \eta_{0}\right)-\ln P\left(s_{1}, \ldots, s_{\tau} \mid \eta_{1}, \ldots, \eta_{\tau}\right)\right] \\
& \approx \sum_{t>0} \underbrace{\mathbb{E}_{Q}\left[\ln Q_{\mu}\left(\eta_{t} \mid a\right)-\ln P\left(\eta_{t} \mid \eta_{0}\right)\right]}_{\text {Risk }} \underbrace{-\mathbb{E}_{Q}\left[\ln P\left(s_{t} \mid \eta_{t}\right)\right]}_{\text {Ambiguity }} \\
\boldsymbol{\mu} &=\arg \min _{\mu} F(s, a, \mu) \\
F&=\sum_{t<\tau} \underbrace{\mathbb{E}_{Q}\left[\ln Q_{ \mu}\left(\eta_{t} \mid a\right)-\ln P\left(\eta_{t+1} \mid \eta_{t}, a\right)-\ln P\left(\eta_{t} \mid \eta_{t-1}, a\right)\right]}_{\text {Complexity }}-\sum_{t \leq 0} \underbrace{\mathbb{E}_{Q}\left[\ln P\left(s_{t} \mid \eta_{t}\right)\right]}_{\text {Accuracy }}.
\end{aligned}
\end{equation}
The conditional independencies among states implicit in partially observed Markov decision processes entail the above functional forms for variational and expected free energies~\cite{dacostaActiveInferenceDiscrete2020,fristonActiveInferenceProcess2017}. Crucially, the posterior over external states uses a mean-field approximation, in which the joint distribution over current and future states factorises into marginal distributions at each point in time [this approximation can be finessed by conditioning on previous states, leading to a different (Bethe) variational free energy~\cite{yedidiaConstructingFreeEnergyApproximations2005,parrNeuronalMessagePassing2019}]. Note that the discrete version of variational free energy is a functional of a distribution over a sequence of states and can be regarded as the discrete homologue of the variational free energy of generalised states in \eqref{eq: free energy generalised}.

The ensuing minimisation of free energy can be formulated as gradient flows following \eqref{eq: 17}—between the discrete arrival of new sensory input—in a way that relates comfortably to neuronal dynamics~\cite{fristonActiveInferenceProcess2017,dacostaActiveInferenceDiscrete2020,dacostaNeuralDynamicsActive2021}. In some simulations, one can mix discrete and continuous state-space models by placing the former on top of the latter, to produce deep generative models that, through active inference, can be used to simulate many known aspects of computational anatomy and physiology in the brain~\cite{fristonGraphicalBrainBelief2017}.

\subsection{Summary}
In summary, we now have at hand a way of identifying the most likely autonomous trajectory from any initial particular state that can be used to simulate the sentient behaviour of precise particles that we have associated with biotic systems. The expected free energy absorbs two aspects of Bayes optimal behaviour into the same (objective) functional~\cite{sajidActiveInferenceBayesian2022}. On a Bayesian reading, the expected information gain is exactly the same quantity that underwrites the principles of optimal Bayesian design~\cite{lindleyMeasureInformationProvided1956,mackayInformationTheoryInference2003,baliettiOptimalDesignExperiments2021}. In other words, the principles that prescribe the best way to solicit evidence that reduces uncertainty about various hypotheses. The second imperative comes from Bayesian decision theory, where the objective is to minimise some expected cost function expected under a choice or decision~\cite{waldEssentiallyCompleteClass1947,brownCompleteClassTheorem1981,bergerStatisticalDecisionTheory1985}. 

Teleologically, it is worth reflecting upon the differences between the generative models that underwrite state-wise and path-wise descriptions of Bayesian mechanics, respectively. For the state-wise formulation \eqref{eq: 21}, the generative model is just a joint density over external and particular states, supplied by—or supplying—the NESS density. For the path-wise formulation \eqref{eq: 27}, \eqref{eq: 34}, the generative model is a joint distribution over the paths of external and sensory states. In other words, there is an implicit state-space model of dynamics that can be summarised heuristically as modelling the consequences of an action on external and sensory dynamics. Because consequences follow causes, the generative model acquires a temporal depth~\cite{yildizBirdsongHumanSpeech2013,fristonDeepTemporalModels2018}. This depth required to describe any given particle may, of course, be another characteristic that distinguishes different kinds of particles. In short, the path-wise formulation describes \textit{particles that plan}, under a proximal or distal horizon.

\section{Conclusion}

There are many points of contact between the variational formulation above and other normative theories of self-organisation and purposeful behaviour. However, to focus the narrative we have deliberately suppressed demonstrating precedents, variants and special cases. Figure \ref{fig: 3} highlights a few relationships between the free energy principle and various formulations of self-organisation and sentient behaviour. In brief, this casts things like reinforcement learning and optimal control theory as optimising the marginal likelihood of particular states, conditioned upon a generative model supplied by a nonequilibrium steady-state density. It could be argued that the link between the free energy principle and established formulations is most direct for synergetics~\cite{hakenSynergeticsIntroductionNonequilibrium1978,hakenInformationSelforganizationUnifying2016} and related treatments of dissipative structures~\cite{prigogineTimeStructureFluctuations1978}. There is also a formal and direct link to information theoretic formulations and Bayesian statistics. Furthermore, the free energy principle can be regarded as dual to the constrained maximum entropy principle \cite{sakthivadivelEntropyMaximisingDiffusionsSatisfy2023}, where the constraints are supplied by the generative model. Please see~\cite{hafnerActionPerceptionDivergence2020,fristonSophisticatedInference2021} for a treatment of things like empowerment~\cite{klyubinEmpowermentUniversalAgentcentric2005}, information bottleneck~\cite{tishbyInformationBottleneckMethod2000} and predictive information~\cite{stillInformationtheoreticApproachCuriositydriven2012,stillThermodynamicsPrediction2012}.

In a similar vein, there are several accounts of optimal behaviour—in both its epistemic and pragmatic aspects—that are closely related to the path integral formulation of active inference. Some key relationships are highlighted in Figure \ref{fig: 7}, such as intrinsic motivation, artificial curiosity~\cite{oudeyerWhatIntrinsicMotivation2007,schmidhuberFormalTheoryCreativity2010,bartoNoveltySurprise2013} and optimal control~\cite{kappenPathIntegralsSymmetry2005,vandenbroekRiskSensitivePath2010,kappenOptimalControlGraphical2012}. The interesting thing about these other theories is that they are predicated on optimising some objective function that can be recovered from expected free energy by taking various sources of uncertainty off the table. This discloses things like the objective optimised in reinforcement learning and expected utility theory in behavioural psychology and economics, respectively~\cite{botvinickHierarchicallyOrganizedBehavior2009,bossaertsBehaviouralEconomicsNeuroeconomics2015a}. 

This paper has focused on a single particle and has largely ignored the (external) context that leads to generalised synchrony among internal and external states. This synchronisation goes hand-in-hand with existence \textit{per se} and the Bayesian mechanics supplied by the free energy principle. The very fact that this mechanics rests upon synchronisation may speak to the emergence of synchronisation among formally similar particles; namely, populations or ensembles. In other words, an individual member of an ensemble or ecosystem owes its existence to the ensemble of which it is a member—at the level of multicellular organisation or indeed its conspecifics in evolutionary biology~\cite{manickaModelingSomaticComputation2019}. In a similar vein, the context established by supra- and subordinate scales plays an existential role. In brief, particles at one scale can only exist if there is a nonequilibrium steady-state density at a higher scale that entails Markov blankets of Markov blankets~\cite{palaciosMarkovBlanketsHierarchical2020}. Due to a separation of temporal scales, much of the self-evidencing at one scale is absorbed into the fast, random fluctuations at the scale above. For example, the fast electrophysiological fluctuations of a neuron become, random fluctuations from the point of view of neuronal population dynamics and sensory motor coordination in the brain~\cite{kelsoDynamicPatternsSelfOrganization1995,decoDynamicBrainSpiking2008,kelsoUnifyingLargeSmallScale2021}. This follows in a straightforward way from applying the apparatus of the renormalisation group. Please see~\cite{fristonFreeEnergyPrinciple2019a} for further discussion.

For brevity and focus, we have not considered applications of the free energy principle and active inference in detail. A brief review of the literature in this area will show that that the majority of applications are in the neurosciences~\cite{dacostaActiveInferenceDiscrete2020} with some exceptions: e.g.,~\cite{fristonKnowingOnePlace2015,ramsteadAnsweringSchrodingerQuestion2018}. Recently, there has been an increasing focus on active inference in the setting of machine learning and artificial intelligence~\cite{ueltzhofferDeepActiveInference2018,tschantzScalingActiveInference2019,barpGeometricMethodsSampling2022,fountasDeepActiveInference2020,hafnerActionPerceptionDivergence2020,catalRobotNavigationHierarchical2021,dacostaRewardMaximizationDiscrete2023}. Much of this literature deals with simulation and modelling and, specifically, scaling active inference to real-world problems. These developments speak to the shift in focus from the foundational issues addressed in this article to their applications. It is quite possible that the foundational aspects of the free energy principle may also shift as simpler interpretations and perspectives reveal themselves.

\section*{Additional Information}

\subsection*{Funding Statement}
KF is supported by funding for the Wellcome Centre for Human Neuroimaging (Ref: 205103/Z/16/Z) and a Canada-UK Artificial Intelligence Initiative (Ref: ES/T01279X/1). L.D. is supported by the Fonds National de la Recherche, Luxembourg (Project code: 13568875). N.S. is funded by Medical Research Council (MR/S502522/1). C.H. is supported by the U.S. Office of Naval Research (N00014-19-1-2556). K.U. was supported by the PRIME programme of the German Academic Exchange Service (DAAD) with funds from the German Federal Ministry of Education and Research (BMBF). This publication is based on work partially supported by the EPSRC Centre for Doctoral Training in Mathematics of Random Systems: Analysis, Modelling and Simulation (EP/S023925/1). The work of GAP was partially funded by the EPSRC, grant number EP/P031587/1, and by JPMorgan Chase \& Co. Any views or opinions expressed herein are solely those of the authors listed, and may differ from the views and opinions expressed by JPMorgan Chase \& Co. or its affiliates. This material is not a product of the Research Department of J.P. Morgan Securities LLC. This material does not constitute a solicitation or offer in any jurisdiction.

\subsection*{Acknowledgements}
We would like to thank Maxwell Ramstead—and participants in his International Physics Reading Group—who went through \cite{fristonFreeEnergyPrinciple2019a} in the forensic detail, generating many of the issues and questions addressed in this paper. We thank our reviewer for providing detailed and helpful feedback that greatly improved the manuscript.

\subsection*{Competing Interests}
The authors have no competing interests.

\subsection*{Authors' Contributions}
All authors made substantial contributions to conception and design, and writing of the article; and approved publication of the final version.



\bibliographystyle{elsarticle-num} 
\bibliography{bib}





\end{document}